\pdfoutput = 1

\documentclass[aps,pre,twocolumn,showpacs,amssymb]{revtex4}

\usepackage{graphicx}
\usepackage{dcolumn}
\usepackage{bm}
\def\be{\begin{equation}}
\def\en{\end{equation}}

\newcommand{\av}[1]{\langle{#1}\rangle}
\newcommand{\Av}[1]{{\bigg\langle}{#1}{\bigg\rangle}}
\def\gs{\gtrsim}
\def\ls{\lesssim}
\newcommand{\bi}[1]{\mbox{\boldmath$#1$}}

\def\bea{\begin{eqnarray}}
\def\ena{\end{eqnarray}}
\def\a{_{\alpha\beta}}

\renewcommand{\theequation}{\arabic{section}.\arabic{equation}}

\begin{document}


\title{ Relationship between 
bond-breakage correlations  
and four-point  correlations in 
heterogeneous glassy dynamics: Configuration changes and  
vibration modes}



\affiliation{Department of Physics, Kyoto University, Kyoto 606-8502,
Japan}


\author{Hayato Shiba$^{1}$\footnote{
These two authors  contributed equally to this work.}, Takeshi  Kawasaki$^{2,*}$, and Akira Onuki$^2$}

\affiliation{$^1$Institute for Solid State Physics, University of Tokyo, 
Chiba 277-8581, Japan\\
$^2$Department of Physics, Kyoto University, Kyoto 606-8502,
Japan}


\date{\today}

\begin{abstract}
We investigate the dynamic heterogeneities of glassy particle systems 
in  the theoretical schemes  of bond breakage and 
four-point correlation functions.  
In  the bond-breakage scheme,  we 
 introduce the  structure factor $S_b(q,t)$ 
and the   susceptibility    
$\chi_b(t)$  
to  detect the spatial correlations of  
configuration changes.  
Here $\chi_b(t)$ attains a maximum 
at $t =t_b^{\rm max}$ as a function of time $t$,  
where the fraction of the particles with broken bonds 
$\phi_b(t)$ is about $1/2$. 
In the four-point scheme,  treating  
the structure factor  $S_4(q,t)$ 
and the  susceptibility   $\chi_4(t)$,   
we    detect  superpositions  of the  
heterogeneity of bond breakage 
and that of   thermal low-frequency vibration modes.  
While the  former grows  slowly,  
the latter emerges  quickly to    exhibit  
complex space-time behavior. In two dimensions, the vibration modes 
extending over the system yield significant contributions to the four-point 
correlations, which depend on the system size logarithmically.   
A  maximum of    $\chi_4(t)$ is attained 
at $t= t_4^{\rm max}$,  where   these two contributions 
become  of the same order. 
As a result, $t_4^{\rm max}$ is  considerably  shorter than 
$t_b^{\rm max}$.  
\end{abstract}



\maketitle


\section{Introduction}
\setcounter{equation}{0}

 Recently, much attention has been paid to the dynamics of glasses  
\cite{Binder}. In particular, 
 dynamic heterogeneities   exceeding   the molecular size and 
emerging  on long timescales  \cite{Adam}  have been  observed  
 in  a number of experiments \cite{Silescu,Ediger,book} and 
molecular dynamics  simulations  in two dimensions (2D) 
and in three dimensions (3D) 
\cite{book,Ha,Eg,Takeuchi,Hi,Hi1,Harrowell,yo,yo1,Kob,Biroli,Dol,Kawasaki,Da,Pe,Ch,Ar}. 
In simulations, they can be detected  if  the spatial correlations 
of the particle configuration changes or 
the displacements between  two separated times are 
calculated.  In an early period,  
 displacement heterogeneities  were observed 
in   applied  strain in model  amorphous alloys \cite{Takeuchi,Eg,Ar}. 
 Harrowell and coworkers  
visualized them  in  a one-component  fluid \cite{Harrowell} 
and  a  binary mixture \cite{Ha}. 
Muranaka and Hiwatari   detected them  
on   short   \cite{Hi} and  long  \cite{Hi1} timescales 
in   binary mixtures. 
 Yamamoto and one of 
the present authors \cite{yo,yo1,yo-diffusion}  
examined breakage of appropriately 
defined bonds and identified 
 relatively active and inactive regions 
without and with applied shear flow. 
The bond-breakage  events are   
produced by  the configuration changes of the particle positions.
The broken  bonds accumulated in long  time intervals 
are heterogeneous such that their  
structure factor $S_b(q,t)$ may be fitted to 
the Ornstein-Zernike form ($\propto 1/[1+q^2\xi_b(t)^2]$), 
where $t$ is the interval width taken to be 
of the order of the structural 
relaxation time  $\tau_\alpha$.
The  correlation length $\xi_b(t)$     grows  with 
lowering  the temperature $T$.   Kob  {\it et al.} \cite{Kob} 
detected  string-like  motions  of mobile 
particles  as fundamental elements   
of structural relaxations, whose length distribution is 
widened with  lowering  $T$.

  La\v{ c}evi\'{c} {\it et al.}\cite{Lacevic} presented 
a statistical  theory of  
the dynamic heterogeneity   in terms of  the so-called 
four-point dynamic correlation functions. 
They  found that the    four-point structure factor $S_4(q,t)$  can  
be fitted to the  Ornstein-Zernike form  
and the  susceptibility   $\chi_4(t)$  exhibits a peak 
at a characteristic time $t_4^{\rm max}$ of order $\tau_\alpha$.  
The  correlation length $\xi_4= \xi_4(t_4^{\rm max})$ thus obtained 
grows with  lowering  $T$.  Subsequently, 
 intensive efforts have been made  to 
construct  statistical theories   
and/or   add  further  numerical results   
 on the four-point correlations 
\cite{book,Sas,b1,b2,Szamel,Chandler,Mizuno,Furukawa}.

However, there has been no systematic comparison between the 
bond-breakage scheme and the four-point scheme. 
The bond-breakage events  
occur  as rare  activation processes, 
resulting in  
 structural relaxations, 
 in the absence of  applied shear. 
In contrast, the physical processes 
giving rise to  the four-point correlations 
have not yet  been well understood. 
 In this paper, we   show that 
the four-point correlations   originate twofold  from  
the configuration changes yielding the bond-breakage correlations 
and from the collective particle motions arising from the  
low-frequency transverse vibration modes 
\cite{Sc,La,Angela,Ganter,Shintani,Bonn,Reichman,Ruocco,Barrat,Liu,Br}. 
The timescales of these two kinds of motions  are dramatically different. 
In the latter,  clusters of relatively mobile  particles 
carry     a large fraction of the vibrational energy  
 and are distributed   throughout  the   system  
\cite{Ruocco}. 
The vibration modes 
have  been studied to explain 
 the low-temperature   thermodynamic 
properties of glasses\cite{Binder}.

In the low-frequency vibrational motions, the oscillatory 
particle   displacements 
 are  highly heterogeneous so that 
the configuration changes   should occur 
preferentially  in more active regions with larger  displacements,  
as pointed out by Schober {\it et al.} \cite{Sc}. 
This   structural relaxation mechanism   
was  confirmed  numerically  in systems with particle  numbers 
about 1000  \cite{Reichman,Br} 
and  experimentally in  quasi-2D colloidal glasses  
\cite{Yodh}. Thus,   it  explains the inseparable coupling 
between  the structural disorder and the slow dynamics in glass.  
We mention some simulations related to this  coupling. 
Vollmayr-Lee {\it et al} \cite{Kob-Binder}  found 
in 3D that mobile particles (in their definition)
are surrounded by fewer neighbors than the others. 
Widmer-Cooper and Harrowell  \cite{Widmer} 
detected a  correlation between
the short-time heterogeneity in a local Debye-Waller
factor and the long-time  heterogeneity in 2D. 
Kawasaki {\it et al.} \cite{Kawasaki} claimed that  
 medium-range crystalline  order remaining in glass 
controls   ease of 
vitrification and nature of the glass transition. 
 In polycrystal with small grains, the relation  
between the structure and the slow dynamics 
is more understandable, 
where the particles at  the grain boundaries  
 initiate   configuration changes \cite{Hama,Shiba,Jack}.

As a closely related effect, 
 a very small applied strain 
 produces  strongly non-affine particle displacements in glass, 
indicating  highly heterogeneous  elastic moduli 
 \cite{Takeuchi,Eg,Ar,Yoshimoto,Barrat-small}. 
Naturally, the particles in 
such elastically  softer  regions exhibit larger-amplitude 
displacements in   the thermally excited 
 vibration modes.  
Tanguy {\it et al.}\cite{Barrat}  showed 
 that the classical  elasticity theory 
 holds only on  spatial scales longer 
 than a characteristic length ($\sim$30 molecular sizes 
in their 2D model system).  Moreover, in glass,  irreversible   
plastic events are induced 
even by very small stains  \cite{yo1,Barrat-small} 
and    plastic deformations  
are highly heterogeneous   often leading  to 
shear bands \cite{book,shear,Shiba}. 
Under a fixed small strain at  $T=0$ in 2D, 
Manning and Liu \cite{soft-spot} numerically examined 
the relation between 
the low-frequency vibration  modes  and 
structural soft spots 
where configuration changes (particle rearrangements in their 
paper) are initiated.

The organization of this paper is as follows.
In Sec.II, our simulation method will be explained. 
 In Sec.III, the bond-breakage scheme  \cite{yo,yo1} will 
be generalized. In Sec.IV, we will 
 reexamine the four-point scheme \cite{Lacevic}, where the collective 
particle motions arising from the vibartion modes will be identified. 
In Sec.V, the dynamic heterogeneities detected by 
 these two schemes will be compared. 
In Sec.VI, 3D  results will be 
presented.

\setcounter{equation}{0}
\section{Numerical method}

To illustrate consequences of the bond-breakage and four-point theories,
 we will show 
results of molecular dynamics simulation 
of $50:50$ binary mixtures composed of 
two   species, 1 and 2,  
 in 2D and 3D in  amorphous states   
 at   low temperatures. 
 We imposed   the periodic boundary condition  
without applying shear flow. 
The particle numbers of the two species are $N_1=N_2=N/2$. 
In  2D,   $N$ will be   mostly 
$4000$ or $64000$, but  data for $N=16000$  and 256000 
will  also be given in Figs.3 and 8. 
In 3D, results for   $N=10000$ 
 will be presented    in Sec.VI. 
  The two species  have different diameters 
 $\sigma_1$ and $\sigma_2$ with 
$\sigma_2/\sigma_1=1.4$ in 2D and 
$\sigma_2/\sigma_1=1.2$ in 3D.  
The  particles   interact via   
the soft-core  potential,  
\begin{equation}
v_{\alpha\beta} (r) = 
\epsilon  \left(\frac{\sigma_{\alpha\beta}}{r}\right)^{12} -C_{\alpha\beta}
\quad  ~
(r<r_{\rm cut}),  
\label{eq:LJP}
\end{equation}
where  $\alpha$ and $\beta$ represent 
the particle species   $(\alpha,\beta =1,2)$, 
 $r$ is the particle  distance, and $\epsilon$  
 is the characteristic interaction energy. 
The  interaction lengths are defined by 
\be     
\sigma{\a} = 
(\sigma_\alpha +\sigma_\beta )/2. 
\en 
The potential vanishes  for   $r>r_{\rm cut}$, 
where $r_{\rm cut}= 4.5\sigma_1$ in 2D  and 
 $r_{\rm cut}= 3\sigma_1$ in 3D. The constants   
 $C_{\alpha\beta}$  ensure the continuity of the potential at 
$r=r_{\rm cut}$.   
 The masses   of the two species satisfy 
  $m_2/m_1= (\sigma_2/\sigma_1)^2$. 
The  average number  density is  
$ 
n=N/V=0.811\sigma_1^{-2} 
$ in 2D and $0.8\sigma_1^{-3}$ in 3D, where   $V$ is the system volume. 
The system length $L$ is $70.2\sigma_1$  for $N=4000$  and 
$281\sigma_1$  for $N=64000$  in 2D, while 
  $L=23.2\sigma_1$ in 3D. 
Space and time will be measured in units of 
$\sigma_1$ and  
\be 
\tau_0= \sigma_1\sqrt{m_1/\epsilon}.  
\en 
The temperature $T$ will be measured 
 in units of $\epsilon/k_B$.

We started from a liquid state 
at a high temperature, 
quenched the system to the final low temperature, and 
waited  for a long time of order  $10^5$. 
  We imposed a thermostat in these steps. 
However, after this preparation of the initial states, we  removed 
 the artificial thermostat and  integrated   the Newton equations 
under the periodic boundary condition 
in the time range $t>0$.  
This is needed  
to describe the effect of 
the vibration modes on long timescales. See the item (3) 
in the summary section for more discussions on the heat bath effect. 
Thus, the total particle number $N$, the total volume $V$, 
and the total energy $E$ are fixed in our simulation.

\setcounter{equation}{0}
 \section{Bond-breakage theory}
\begin{figure}
\includegraphics[width=0.860\linewidth, bb = 0 0 533 696]{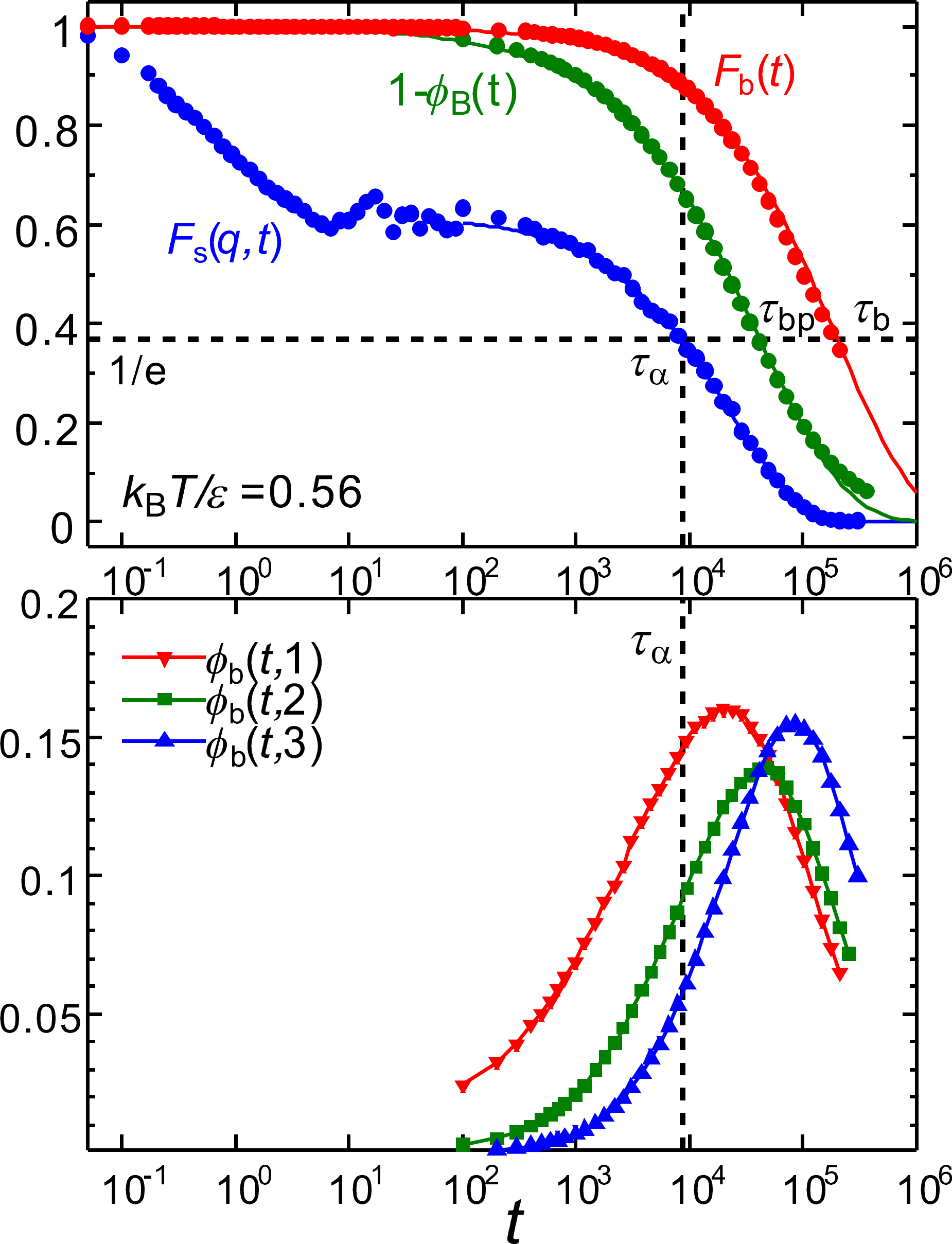}
\caption{
Top: Bond relaxation function 
$F_b(t)$ in Eq.(3.3),  
self part of density time-correlation function 
 $F_s(q,t)$ at $q=2\pi$  
in Eq.(3.5), and  
fraction of non-${\bi B}$ particles 
$1-\phi_b(t)= \phi_B(t,0)$ in Eq.(3.15)  
 at $T=0.56$ for $N=4000$ in 2D.
Relaxation times here  are 
$\tau_\alpha=8400$ from Eq.(3.6), 
$\tau_b= 2.0\times 10^5\cong 35 \tau_\alpha$ from Eq.(3.4). 
and $\tau_{bp}=4.18\times 10^4\cong 5\tau_\alpha$ from Eq.(3.16). 
Bottom: fractions of ${\bi B}$ particles 
with $k$ broken bonds  $\phi_B(t,k)$ in Eq.(3.14) for $k=1,2,$ and 3. 
}
\end{figure}

\begin{figure}[t]
\includegraphics[width=0.80\linewidth, bb=0 0 497 374]{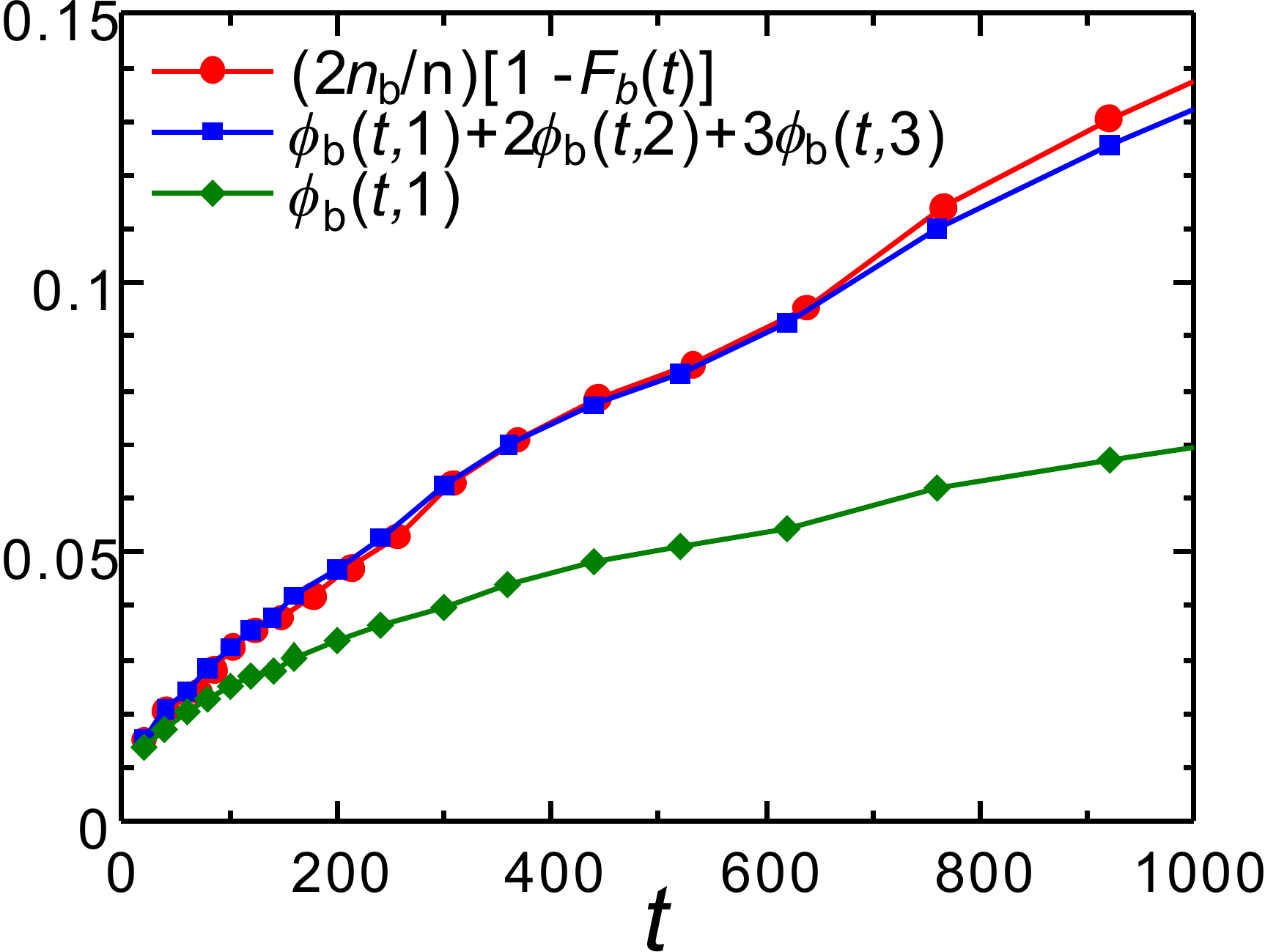}
\caption{Numerical results of 
$(2n_b/n)[1-F_b(t)]$, 
$\phi_B(t,1)$, and $\phi_B(t,1)+2\phi_B(t,2) 
+3\phi_B(t,3)$ as functions of  $t$ 
in the early stage at $T=0.56$ for $N=4000$,  
 which    confirm  Eq.(3.18). 
}
\end{figure}

\subsection{Background}
 
We regard    two particles $i$ and $j$ 
with positions ${\bi r}_i (t)$ and ${\bi r}_j (t)$ 
to be bonded if \cite{yo,yo1}
\be 
r_{ij}(t)<A_1\sigma\a,
\en   
where $i \in \alpha$ and $j \in \beta$. 
Hereafter 
$r_{ij}(t)= |{\bi r}_i (t)-{\bi r}_j (t)|$ is the 
distance between these particles 
at time $t$. At a later time $t+\Delta t$, 
this  bond is   treated to be  broken if 
\be 
r_{ij}(t+\Delta t)>A_2 \sigma\a.
\en  
 We assume that   
$A_1\sigma_{\alpha\beta}$ 
is slightly larger than the peak distance 
of the pair correlation 
functions $g_{\alpha\beta}(r)$ 
and  $A_2$ 
is somewhat larger than $A_1$.
In this paper, we set $A_1= 1.15$ and $A_2=1.5$ in 2D and 
 $A_1= 1.3$ and $A_2=1.7$ in 3D.

Let us consider the bonds  at $t=t_0$ and denote 
their total number as $N_{b}(t_0)$.   
A fraction of them will be broken subsequently 
 and the total number of the remaining bonds 
$N_{b}(t_0+\Delta t)$  at  $t=t_0+\Delta t$ 
decays  as 
\be 
N_{b}(t_0+\Delta t)/N_{b}(t_0)= F_{b}(\Delta t).
\en  
For  large systems, the relaxation function  
 $F_b(\Delta t)$ may be treated to depend  only on the 
 time difference $\Delta t$ (being nearly independent 
 of the initial time $t_0$ for large $N$).  It  decreases 
 with increasing $\Delta t$, so 
  the bond-breakage time  $\tau_b$ may be defined by 
\be 
 F_{b}(\tau_b)= e^{-1}.
\en 
On the other hand, the self part of the density 
 time-correlation function is expressed as 
 \be  
 F_s(q,t)=\frac{1}{N}\Av{\sum_j \exp[
 i{\bi q}\cdot \Delta{\bi r}_j(t_0,t_0+t)]}.
\en 
where 
$\Delta{\bi r}_j(t_0,t_1)= {\bi r}_j(t_1)- {\bi r}_j(t_0)$ 
is the  displacement vector 
of particle $j$ and $\bi q$ is the wave vector. In Eq.(3.5), 
the average  is taken over all the particles. 
In our simulation,  the average  
 over the initial time $t_0$   and that 
over a number of runs were  also taken. 
The structural 
 relaxation time $\tau_\alpha$ is usually  defined  at  $q=2\pi$ by   
\be 
F_s(q,\tau_\alpha)=e^{-1}.
\en 
Wave numbers will be measured  
in units of $\sigma_1^{-1}$.

In the upper panel of Fig.1, 
we display  
$F_{b}(t)$ and $F_s(q,t)$ at  $q=2\pi$  for  $T=0.56$  and $N=4000$, where 
$\tau_\alpha=8.4\times 10^3$ and  
$\tau_b= 2.0\times 10^5 \cong 35 \tau_\alpha$.
In the previous paper \cite{yo1},   
the relation $\tau_b \cong 10 \tau_\alpha $ was found  for  
binary mixtures with the soft-core 
potentential wih $N=10^4$  in 3D.  Both in 2D and 3D, 
  $F_{b}(t)$ may  fairly  be  fitted to 
 the stretched exponential  form at low $T$ as  
 \be 
 F_{b}(t) \cong \exp[-(t/\tau_b)^c], 
\en 
in the range  $t\ls \tau_b$. 
At  $T=0.56$ we obtain  $c \sim 0.58$ in 2D. 
The  exponent $c$ increases   with decreasing $T$. 
In contrast,  
$F_s(q,t)$ exhibits a plateau $f_{\rm p}(<1)$ 
before  the  $\alpha $ relaxation   at  low $T$  due to 
the thermal vibrational  motions  
(see the appendix).

\subsection{Bond-breakage correlations}

Here, we present 
 a generalized  formulation of  
the bond breakage to 
introduce broken-bond  correlation 
functions.  To this end, 
 we define two overlap functions $w\a^{(1)}({ r})$ 
and $w\a^{(2)}({r})$ 
depending on the particle distance $r$ as  
\be 
w\a^{(K)} (r)= \theta ( A_K \sigma\a-r),
\en 
with $A_1$ and $A_2$ being defined in Eqs.(3.1) and (3.2).  
The  $\theta(u)$ is the step function 
being equal to 1 for $u>0$ and to  0 for $u\le 0$. 
The fluctuating number density of the bonds may 
then be defined as   
\be 
\hat{n}_b({\bi r},t)= \frac{1}{2} 
\sum_{ij}w\a^{(1)}(r_{ij}(t))  \delta ({\bi r}- {\bi r}_i(t)), 
\en 
where we multiply  $1/2$  because 
a   bond is supported by two particles in our  definition. 
The  statistical average of  $\hat{n}_b({\bi r},t)$
is the average bond number density,    
\be 
n_b= \frac{1}{V} \int {d{\bi r}}
{\hat{n}_b({\bi r},t)}= \frac{1}{V} N_b(0).  
\en 
Here,   $n_b \sim 3n$  in 2D at high densities.
In fact, we numerically 
obtain  $n_b=2.28=2.81n $ for $n=0.811$ 
in our 2D system.  
Now we may introduce  the broken bond number density 
in  time interval $[t_0,t_1]$  as  
\be
\hat{\cal P}({\bi r},t_0,t_1)=  \frac{1}{2} 
\sum_{i}{\cal B}_i (t_0,t_1) 
 \delta ({\bi r}- {\bi r}_i(t_0)),   
\en 
where   ${\cal B}_i(t_0,t_1)$  is  the broken bond number of particle $i$ 
assuming  a nonnegative integer quantity   as   
\be 
{\cal B}_i (t_0,t_1) 
=  \sum_j  w\a^{(1)}(r_{ij}(t_0)) 
[1-w\a^{(2)}({r}_{ij}(t_1)) ].  
\en  
This number tends to  zero as  $t_1 \to  t_0$ from $A_1<A_2$ 
and increases to $1, 2, \cdots$  upon bond breakage.   
Hereafter, the particles with  
${\cal B}_i (t_0,t_1)\ge 1$ 
are  called  $\bi B$ particles, which  
are  surrounded by different  particle 
configurations at the initial and final times 
$t=t_0$ and $t_1$. 
On the other hand, those with 
${\cal B}_i (t_0,t_1)=0$ are 
called  non-$\bi B$ particles,  which 
have  the same surrounding  configurations  
at  $t=t_0$ and $t_1$. The  statistical  average $ 
\av{\hat{\cal P}({\bi r},t_0,t_1)}$ 
 depends on the time difference $t=t_1-t_0$  as    
\be
p_b(t) = \frac{1}{V} \int {d{\bi r}}
 {\hat{\cal P}({\bi r},t_0,t_1)}= 
 n_b[1-F_b(t)]
\en  
where $n_b$ is defined by Eq.(3.10) 
and $F_b(t)$  by Eq.(3.6).

Let the  number of 
the particles with ${\cal B}_i (t_0,t_1)=k$ 
be $N_{B}(t,k)$ ($k=0,1,\cdots$) with $t=t_1-t_0$. 
Then, 
\be 
\phi_B(t,k)= N_B(t,k)/N
\en 
is the fraction of the ${\bi B}$  particles with 
$k$ broken bonds for $k \ge 1$, while 
$\phi_B(t,0)= N_B(t,0)/N$ is the fraction 
of the  non-$\bi B$ particles.  
The fraction 
of the total $\bi B$ particles is the following sum, 
\be 
\phi_b(t)= \sum_{k \ge 1} \phi_B(t,k)=1-\phi_B(t,0).
\en 
We define  the bond-preserving time $\tau_{bp}$ as 
\be 
1-\phi_b(\tau_{bp})= \phi_B(\tau_{bp},0)=e^{-1}.
\en 
The particles have a broken bond on this time scale. 
Since  each particle has several bonds 
($\sim  6$ in 2D),  $\tau_{bp}$ is 
considerably shorter  than 
the bond breakage time $\tau_b$ in Eq.(3.4). 
For $t\gs \tau_{bp}$, the structural relaxation becomes appreciable. 
From Eq.(3.14) we obtain 
\be 
p_b(t) = \frac{n}{2}  
\sum_{k } k \phi_b(t,k)
\en 
From Eqs.(3.13) and (3.17) we find 
\be 
 2\frac{n_b}{n}  [1- F_b(t)]= 
\sum_{k } k \phi_b(t,k)
\en

Setting $t=t_1-t_0$ in steady states, 
  we introduce the bond-breakage space-time correlation function,
\bea 
G_b({r},t)&=& \av{\hat{\cal {P}}({\bi r}+{\bi r}',t_0,t_1) 
\hat{\cal {P}}({\bi r}',t_0,t_1)} 
\nonumber\\
&&\hspace{-2.5cm} 
=\frac{1}{4V}\Av{
\sum_{ik}  
{\cal B}_i (t_0,t_1) {\cal B}_k (t_0,t_1)  
\delta ({\bi r}- {\bi r}_{ik}(t_0)) },
\ena
where ${\bi r}_{ik}(t_0)
={\bi r}_i(t_0)-{\bi r}_k(t_0)$. 
From Eq.(3.13) we have $G_b(r,t) \to p_b(t)^2$ for large $r$. 
The structure factor 
of the broken bonds is   given by  
\bea
S_b(q,t)&=& \frac{1}{V} \av{|\hat{\cal P}_{\bi q}(t_0,t_1)|^2}\nonumber\\
&&\hspace{-1cm}= \int d{\bi r}[G_b({\bi r},t)-p_b(t)^2]
e^{i{\bi q}\cdot{\bi r}},
\ena
where $\bi q$ is the wave vector, 
$q=|{\bi q}|$ is the wave number, 
and ${\hat{\cal P}}_{\bi q}(t_0,t_1)= 
\sum_{j}{\cal B}_j (t_0,t_1) 
 \exp[i{\bi q}\cdot{\bi r}_i(t_0)]/2$ is the 
 Fourier component of ${\hat{\cal P}}({\bi r}, t_0,t_1)$.

As in the four-point scheme  \cite{Lacevic}, we 
introduce  the  susceptibility $\chi_b(t)$  to  
represent  the overall degree of the bond-breakage correlations  as follows:   
\be 
\chi_b(t)=\frac{1}{4V} \Av{
\sum_{ik} \delta{\cal B}_i (t_0,t_1) \delta{\cal B}_k (t_0,t_1) }, 
\en 
in terms of the deviations  
 $\delta {\cal B}_i (t_0,t_1)= 
{\cal B}_i (t_0,t_1)- 2p_{b}(t)/n$. 
Here $\av{{\cal B}_i}= 
\sum_i {\cal B}_i/N= 2 p_b/n$ from Eqs.(3.11) and (3.13). 
In $S_b(q,t)$ and $\chi_b(t)$, the four particle 
positions ${\bi r}_i(t_0)$, ${\bi r}_k(t_0)$, 
${\bi r}_j(t_1)$, and ${\bi r}_\ell(t_1)$ 
are involved. In this sense, they are four-point 
correlation functions.

Rabini {\it et al.} \cite{Rabini} 
 introduced  the number of particles
 that have left particle $i$'s original neighbors  at  time $t$.  
 It involves two times and 
was written as $n_i^{\rm out}(0,t)$. 
It is similar to our ${\cal B}_i(t_0,t_0+t)$ in Eq.(3.12).  
Abate and Durian \cite{Durian} 
 introduced a  bond-breakage susceptibility 
$\chi_B(t)$ similar to  ours in Eq.(3.21)  
for a quasi-two-dimensional granular system 
of  air-fluidized beads.

\begin{figure}[t]
\includegraphics[width=0.85\linewidth, bb=0 0 590 409]{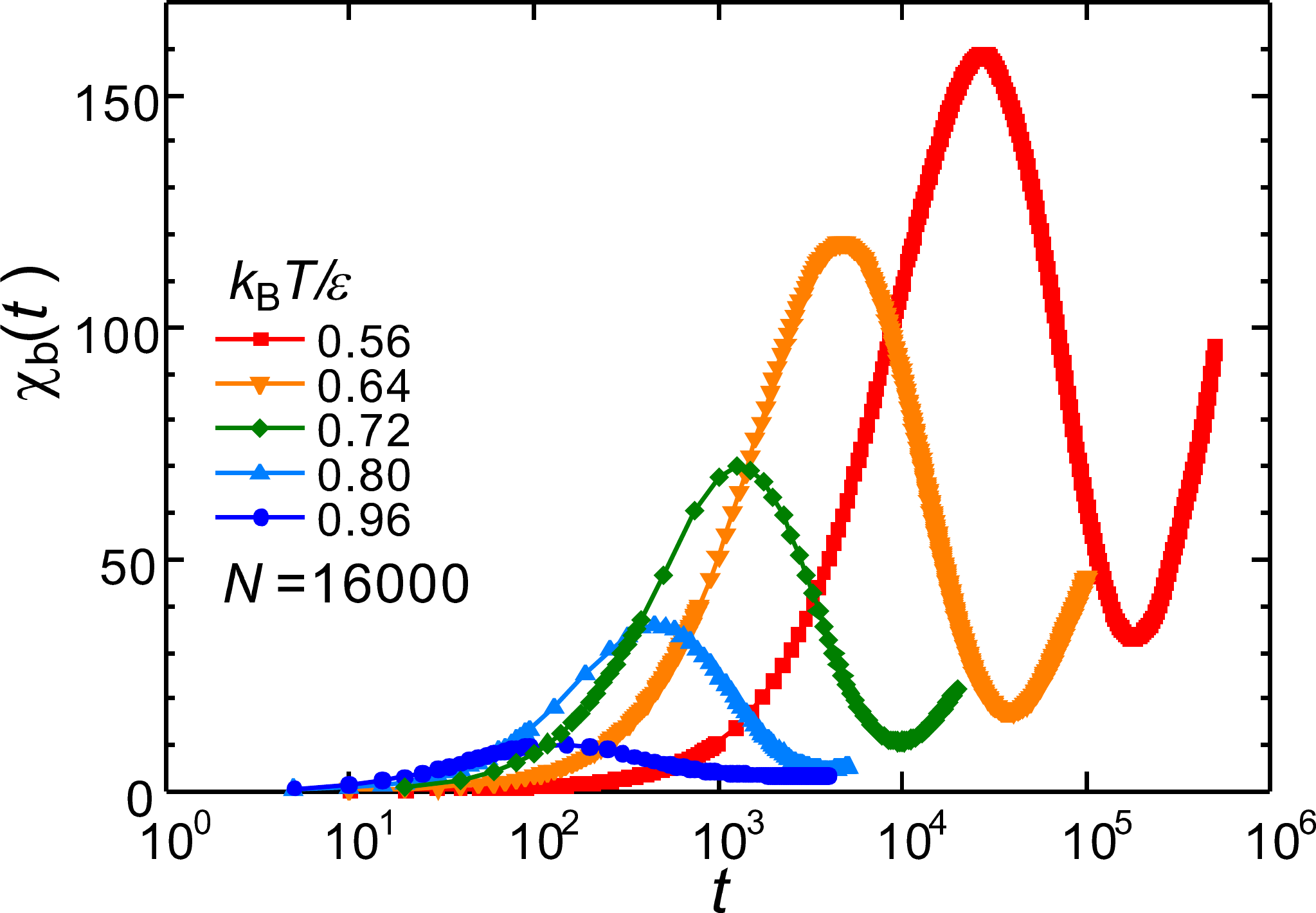}
\caption{ Susceptibility 
 $\chi_b(t)$ in Eq.(3.21) vs $ t$ exhibiting a peak at $t=
t_b^{\rm max}$ for $N=16000$.
}
\end{figure}

\begin{figure}[t]
\includegraphics[width=0.85\linewidth, bb=0 0 487 561]{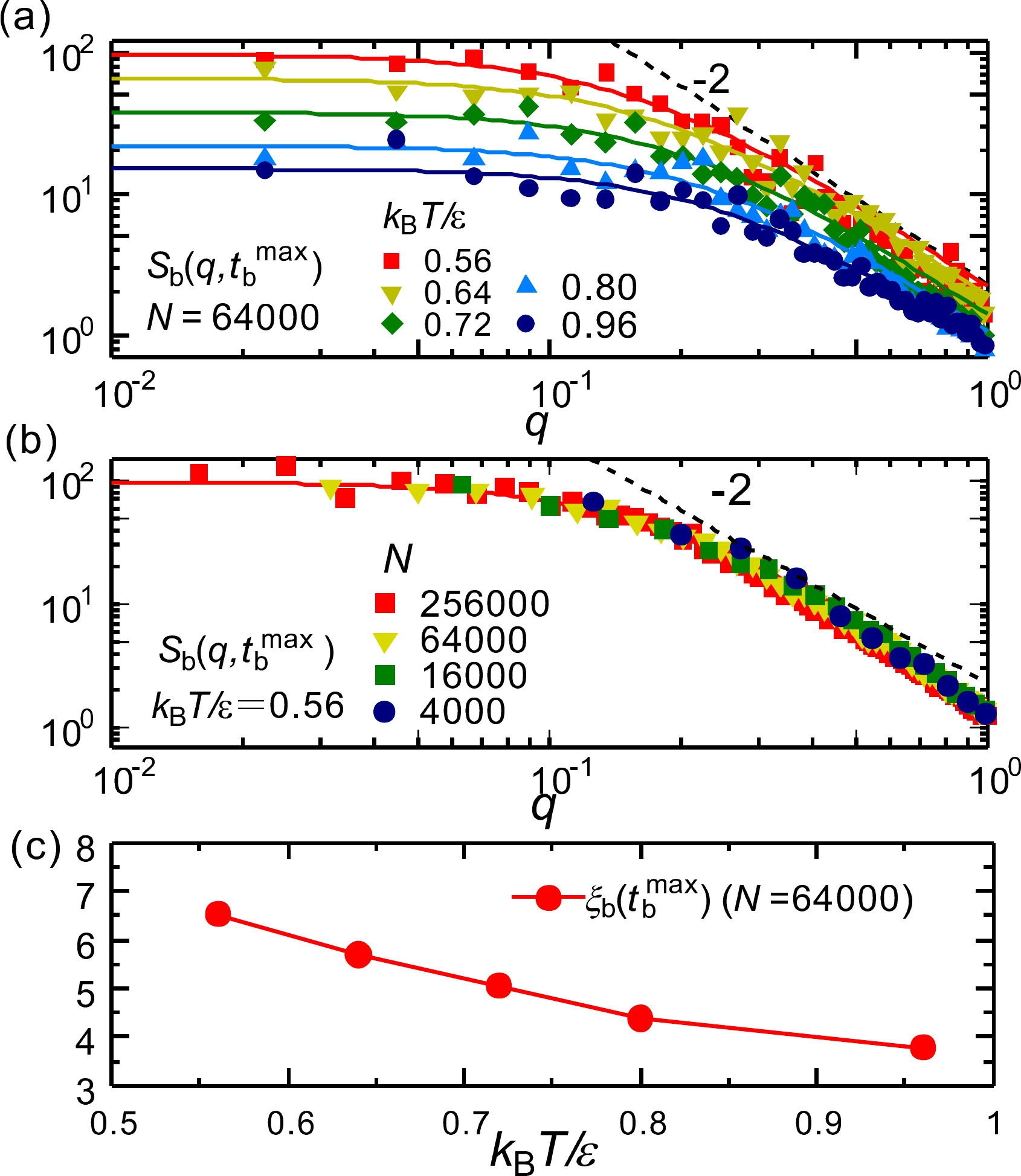}
\caption{(a) Structure factor 
$S_b(q,t_b^{\rm max})$  for bond breakage in Eq.(3.20)  at  various $T$ 
for $N=64000$,    (b) $S_b(q,t_b^{\rm max})$ at $T=0.56$ 
for various $N$, 
and  (c) correlation lengths 
$\xi_b= \xi_b(t_b^{\rm max})$ vs $T$  for $N=64000$.  
$S_b(q,t)$ is approximately on a single curve 
independent of $N$, leading to weak system-size dependence 
of $\xi_b$ 
for  $1\ll \xi_b \ll L$.  
}
\end{figure}

\subsection{Numerical results on bond breakage}

We further discuss consequences 
of our theory using numerical results  in 2D. 
In Fig.1, we plot $1-\phi_b(t)=\phi_B(t,0)$ 
in the upper panel and $\phi_B(t,k)$ with $k=1,2$, and 3 
in the lower panel, where   
$\tau_{bp}=41800\cong 
5\tau_\alpha= 0.14\tau_b$ from Eq.(3.16). 
At small $t$, $\phi_B(t,k)$ 
grow  as   
\be 
{\phi}_B(t,k) \propto t^{a_k},
\en 
where $a_1 \sim 0.60$, $a_2 \sim 1.0$, and $a_3 \sim  1.3$. 
Though we cannot derive these exponents theoretically, they should 
arise from correlated occurrence of bond breakage events.
   
In Fig.2, we confirm the validity of Eq.(3.18) 
from  simulation  at $T=0.56$, where  
 $n_b/n=3.17$.  We find that 
$(2n_b/n)[1-F_b(t)]$ 
nearly coincides with 
$\phi_B(t,1)$ for $t <200$ 
and with $\phi_B(t,1)+2\phi_B(t,2) 
+3\phi_B(t,3)$ for $t<4000$ within a few $\%$ differences.    
The first relation for $k=1$ in Eq.(3.22)  is consistent with 
Eq.(3.18) since $a_1\cong  c$. In fact, for 
$t\ll \tau_b$,    Eqs.(3.7) and (3.18) yield   
\be 
\phi_B(t, 1) \cong (2{n_b}/{n}) (t/\tau_b)^c. 
\en 

In Fig.3, we plot $\chi_b(t)$ in Eq.(3.21) as  a 
function of $t$ for $N=16000$, which exhibits a peak at $t=t_b^{\rm max}$. 
Here,   we have 
$\phi_b(t) \sim 0.5$ at  $t=t_b^{\rm max}$, 
which can be seen in Fig.1 for $N=4000$ and $T=0.56$.
Here, however, $\chi_b(t)$ begins to increase for $t>t_b^{\rm max}$, 
 because the $\bi B$ particles with ${\cal B}_i=2,3,\cdots$ 
becomes appreaciable at very long times (see Fig.1). 
In (a) of  Fig.4, we show    
  $S_b(q,t_b^{\rm max})$ vs $q$ 
  for various $T$ with  $N=64000$. 
In its calculation  we took the average 
over the initial time $t_0$ in a wide range of 
$[0, 10^6]$ for $T\ge 0.64$ and 
that of $[0, 2\times 10^6]$ for $T= 0.56$, 
which was needed because of the slow bond breakage. 
We may   fairly   fit  $S_b(q,t)$ to  the 
 Ornstein-Zernike form \cite{yo,yo1}, 
 \be 
 S_b(q,t) = \chi_b^0(t)  /[1+q^2\xi_b(t)^2], 
 \en 
 where $ \chi_b^0(t) =\lim_{q\to 0} S_b(q,t) $ 
is the long wavelength limit of $S_b(q,t)$ and 
$\xi_b=\xi_b(t)$ is the  correlation  length 
representing  the  spatial scale of the 
correlated configuration changes.  Furthermore, in (b),  
we  show $S_b(q,t_b^{\rm max})$ vs $q$ 
 at $T=0.56$  for various  $N$,  which demonstrates weak 
system-size dependence of the bond-breakage correlations. 
For   $N=256000$, however,  the averaging over the initial time $t_0$ 
is still insufficient because of very large  
$t_b^{\rm max}\sim 4\times 10^4$ as compared to the 
simulation time ($\sim 10^5$). As a result, 
the corresponding  $S_b(q,t_b^{\rm max})$ exhibit noticeable 
fluctuations at small $q$.   In (c), the correlation length  $\xi_b$ 
vs $T$ is plotted at $t=t_b^{\rm max}$ for $N=64000$, 
which increases  with lowering $T$. Note that 
 $\xi_b= \xi_b(t_b^{\rm max})$ 
is nearly independent of the system size from (b) 
 as long as  $1\ll \xi_b\ll L$.   

{

In the original papers\cite{yo,yo1}, 
 the broken-bond  structure factor was defined 
 differently, so it is written  as $S_b^{\rm YO}(q,t)$ here.  
  It was calculated for  
  the Fourier component of 
 the following broken bond number density, 
\bea 
&& \hspace{-1cm}
\hat{\cal P}_{\rm YO}({\bi r},t_0,t_1)= 
  \frac{1}{2} \sum_{ij} 
  w\a^{(1)}(r_{ij}(t_0)) 
[1-w\a^{(2)}({r}_{ij}(t_1)) ]\nonumber\\
&&
 \times \delta ({\bi r}- {\bi R}_{ij}(t_0) ),   
\ena 
where  $t=t_1-t_0$ was 
 set equal to  $0.05 \tau_b$ or $0.1\tau_b$. 
  Here,  the midpoint position  
${\bi R}_{ij}(t_0) =
 \frac{1}{2}({\bi r}_i(t_0)+{\bi r}_j(t_0))$ 
 of the two particles  $i$ and $j$ is  used 
instead of the position ${\bi r}_i(t_0)$ 
in $\delta ({\bi r}- {\bi r}_{i}(t_0) )$ 
in Eq.(3.11).    Note that  the particles 
 pairs with a common broken bond are included 
in $S_b(q,t)$. As a result,  the inter-particle 
 peak at $q\cong 2\pi $ appears in $S_b(q,t)$ 
 (which is not shown in Fig.4), while 
it does not apppear in $S_b^{\rm YO}(q,t)$.  
However,    
there is   no essential difference 
between  these two definitions  for $q< 2$. 
 In addition, in the  previous work \cite{yo1}, 
 the dynamic scaling relation  
of the form $\tau_b \sim \xi_b^z$ was obtained, where 
   $z=4$  in 2D and $z=2$  in 3D.

\section{Four-point theory}
\setcounter{equation}{0}

\begin{figure}
\includegraphics[width=0.90\linewidth, bb = 0 0 532 395]{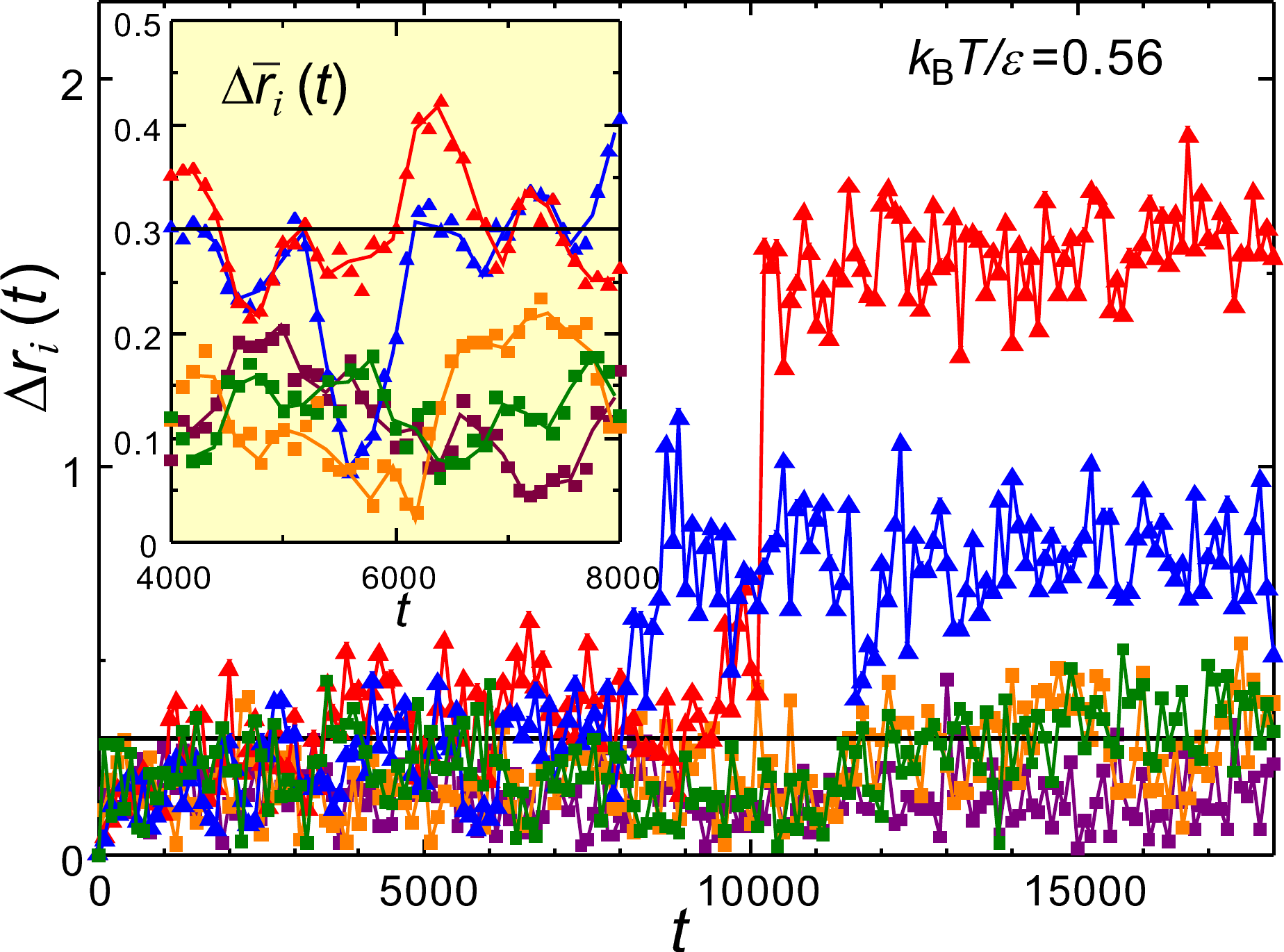}
\caption{
Time evolution of $\Delta r_i(t_0, t_0+t)$  
for five particles among  $4000$ particles. 
They  exhibit  thermal fluctuations 
with magnitudes on the order of the overlap length $0.3$. 
Two  particles  escape  from the initial 
circle at $t \sim 8000$ (blue $\triangle$)
and $\sim 10^4$ (red $\triangle$).  They have  a broken bond 
 with ${\cal B}_i (t_0,t_0+t)  =1$ and ${\cal F}_i (t_0,t_0+t)  =0$  
 in later times.  For the other three particles, 
 the bonds are preserved, but ${\cal F}_i$ 
 frequently fluctuate  between 1 and 0 in the time 
 range displayed. Inset: Time evolution of 
 the time-smoothed distance $\Delta {\bar r}_i(t)$ 
 defined in Eqs.(4.8) and (4.10) in the range $4000<t<8000$.  
 }
\end{figure}

\begin{figure}[t]
\includegraphics[width=0.85\linewidth, bb = 0 0 468 609]{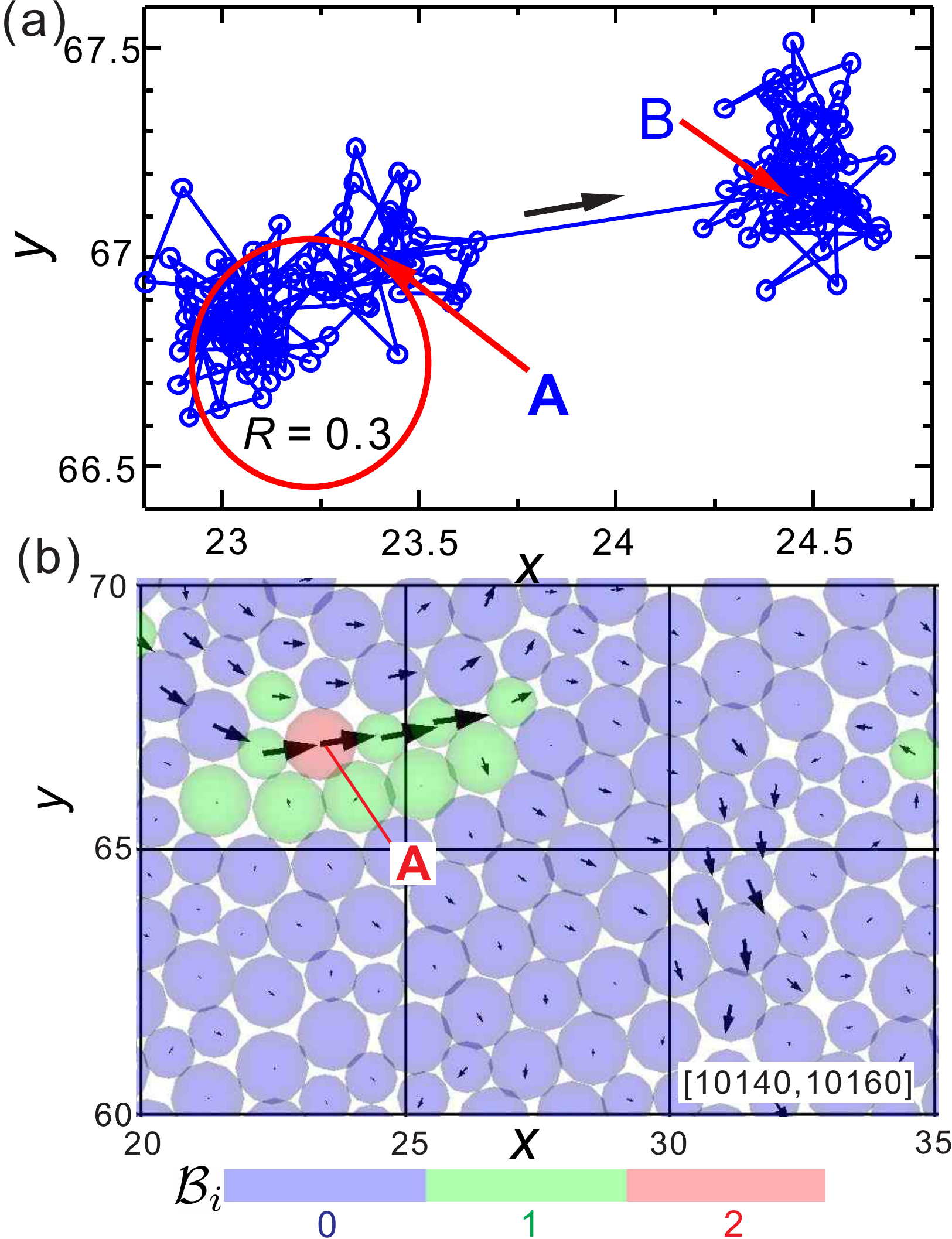}
\caption{(a) Trajectory of the 
particle escaping from a cage 
(red line in Fig.5). Data points 
at $t= 8000+20k$ $(k=0,1,\cdots, 180)$ are written.   
Its position is $(x,y)=(23.36, 67.00)$ at $t= 10140 $ (point A)  and 
$(x,y)= (24,44, 67.16)$  at $t=10160$ (point B). 
Between these times, the particle escapes  from the circle 
 $|{\bi r}- {\bi r}_i (t_0)|<0.3$ (red line). 
(b)  Displacements  between this  short time interval $[10140, 10160]$. 
Several particles 
  undergo a string-like 
motion, where the particle (A) in the upper panel 
has two broken bonds and the others have  one broken bond. 
 }
\end{figure}

La\v{ c}evi\'{c} {\it et al.}\cite{Lacevic}  introduced the 
four-point correlation function to analyze 
the dynamic heterogeneity in glassy systems. 
In their numerical analysis of a 3D  binary mixture in the NVE 
ensemble, they used the 
Lennard-Jones potential, where the particle size ratio was 
  $\sigma_2/\sigma_1=1.2$ and 
the particle numbers were $N_1=N_2= 4000$. 
We  critically  review their theory  
comparing it with our theory  of  bond breakage 
using  some numerical analysis.

\subsection{Overlap   and nonoverlap with  initial regions }

For a time interval $[t_0,t_1]$ ($t=t_1-t_0>0$), 
we introduce  a fluctuating density variable,  
\be 
\hat{\cal{Q}}({\bi r},t_0,t_1)= 
\sum_{i} {\cal F}_i (t_0,t_1) 
\delta ({\bi r}- {\bi r}_i(t_0)) .
\en 
For each $i$   we define  a nonnegative integer, 
\be 
{\cal F}_i (t_0,t_1) = \sum_j 
w(|{\bi r}_i(t_0)-{\bi r}_j(t_1)|) ,
\en 
using  the following overlap function\cite{Lacevic},  
\be 
w(r)= \theta (A_4 \sigma_1 - r).  
\en  
The overlap length  $A_4 \sigma_1$  is common 
 for the two particle species for simplicity. 
 In Eq.(4.2)  the particle positions 
 ${\bi r}_i(t_0)$ and ${\bi  r}_j(t_1)$ are those at different times. 
Thus  ${\cal F}_i (t_0,t_1) $ is 
 the number of overlapping  particles 
 in the initial  circle (or sphere in 3D)  $|{\bi r}-{\bi r}_i(t_0)| 
 < A_4\sigma_1$  in
two configurations separated by time $t= t_1-t_0$. 
We may call $\hat{\cal{Q}}({\bi r},t_0,t_1)$ 
the two-point overlap density.

In the original analysis 
 \cite{Lacevic}, the overlap function was  
written as  $w(r)= \theta (a \sigma_2 -r)$ 
with   $a\sigma_2 =1.2 a\sigma_1$, 
where the parameter $a$ was  set equal to 0.3 
maximizing the four-point 
susceptibility $\chi_4(t)$ 
(see the discussion around  Eq.(4.17)  below).  
 In numerical analysis in this paper, 
 we  set   $A_4=0.3 $ in Eq.(4.3). 
 These  selected values     are  
 considerably shorter  than 
the particle radii, but somewhat exceed the square root of 
the plateau value of the mean square displacement \cite{Lacevic}. 
Thus, as  $t_1 \to t_0$,  
 the distinct terms  with  $j\neq i$ vanish in 
the summation of Eq.(4.2), leading to   ${\cal {F}}_i(t_0,t_1)\to 
1$    and $\hat{\cal{Q}}({\bi r},t_0,t_1)\to 
\hat{\rho}({\bi r},t_0) 
$, where $\hat{\rho}({\bi r},t)=  \sum_{i} \delta ({\bi r}- {\bi r}_i(t)) 
$  is  the usual fluctuating number density.  
For  $t_1>t_0$,    ${\cal {F}}_i(t_0,t_1)$ is  either   0 or 1. 
We did not detect the particles with   ${\cal {F}}_i\ge 2$ 
in our simulation.    

In the definition of  $\hat{\cal{Q}}({\bi r},t_0,t_1)$,  
 the terms in Eqs.(4.1) and (4.2) may be divided 
into the self part with $i=j$ and the distinct part 
with $i\neq j$. The self part of  $\hat{\cal{Q}}$ reads 
\be 
\hat{\cal{Q}}_s({\bi r},t_0,t_1)= 
\sum_{i}  w(\Delta r_i(t_0,t_1))
\delta ({\bi r}- {\bi r}_i(t_0)),   
\en 
where $\Delta r_i(t_0,t_1)$ is 
 the displacement length of particle $i$,   
\be 
\Delta r_i(t_0,t_1)= 
|{\bi r}_i(t_1)-{\bi r}_i(t_0)|. 
\en 
Setting  $a=0.3$, 
 La\v{ c}evi\'{c} {\it et al.}\cite{Lacevic} found that 
  the self part gives rise to the dominant contributions 
in  the four-point correlations.  Also in our simulation, 
the self part dominates over  the distinct part. 
We  consider the particles with 
$\Delta { r}_i>0.3$ and  ${\cal F}_i=1$,  for which  
another  particle $j$ has moved 
within the initial circle of particle $i$ 
giving rise to the distinct contribution.  
 We compare their number  with 
the  number of the  particles with $\Delta { r}_i>0.3$.  
For example, in a 2D  simulation run   for $T=0.56$ 
 and  $N=4000$ 
(which yielded  Fig.13 and the left panels of 14), 
these numbers are  256  and 2195, respectively, 
 at  $t=10200$ (see Table II). 
 In a 3D  simulation run   for 
$T=0.56$  and  $N=4000$ (which yielded  Figs.15 and 16), 
they are  366 and 1452, respectively,  at $t=10^4$ (see Table IV). 
More than $90\%$  of these 
distinct particles (with $\Delta { r}_i>0.3$ and  ${\cal F}_i=1$)   are  
 $\bi B$ particles  having  broken bonds. 

\subsection{Background vibrational  fluctuations}

\begin{figure}[t]
\includegraphics[width=0.85\linewidth, bb= 0 0 347 185]{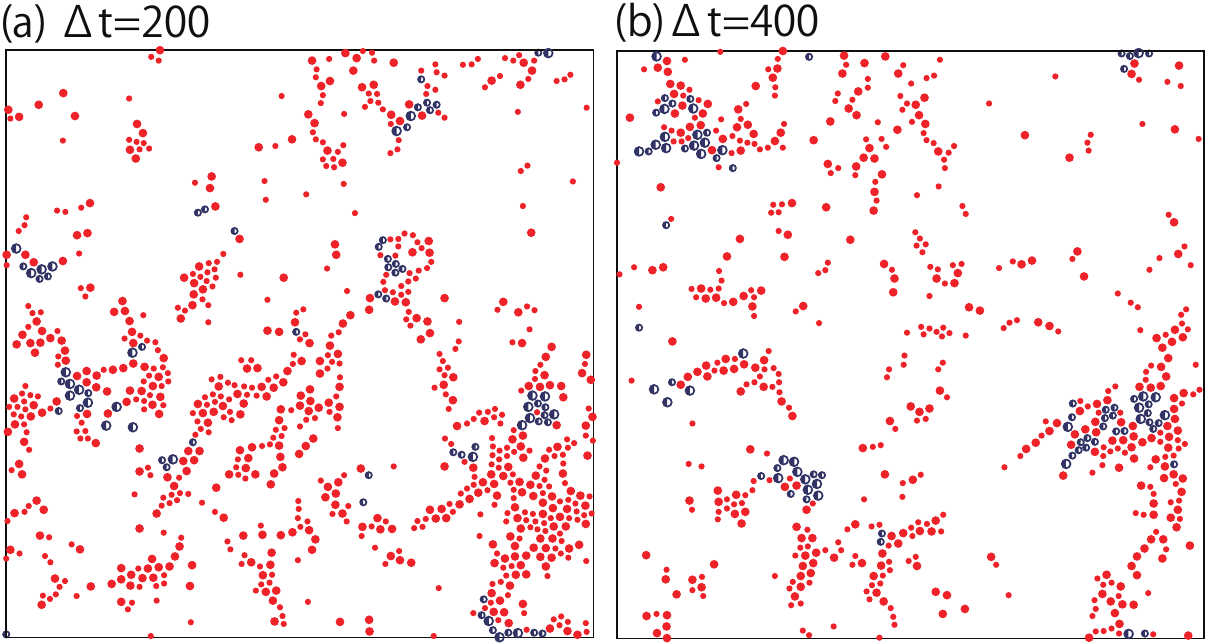}
\caption{Early-stage snapshots 
of particles with  ${\cal F}_i=0$ 
 at $t=200$  (left) and 400  (right), 
which are classified into those with 
 ${\cal B}_i=0$ (red) 
and those with   ${\cal B}_i=1$ (blue).  
Here   $T=0.56$ and 
$N=4000$.   Numbers 
of the former (red) and  the latter (blue)  
are $(766,75)$ in (a) and $(535,81)$ in (b). 
 The former arise from the collective 
vibrational modes emerging for $t\gs 10$. 
}
\end{figure} 

We  first examine the  fluctuations of the  particle 
positions due to  the thermally  excited low-frequency vibration modes 
at low $T$ before  the  onset of the structural  relaxation. 
They exhibit  significant  heterogeneity,   
as found by  Muranaka and Hiwatari  
 in  a very short time interval of width 5 
 in a 2D soft-core system with  $N=10^4$  \cite{Hi}.

 For $ t\ll \tau_{bp}$,  we may neglect the configuration changes 
   from the discussion below Eq.(3.16) 
and define  the dispalcement vector 
${\bi u}_i(t)= {\bi r}_i(t)-{\bar{\bi r}}_i$, 
where ${\bar{\bi r}}_i$ is the time-averaged position 
in a time interval with width much shorter than $\tau_{bp}$. 
The  equal-time variance $\av{|{\bi u}|^2}=\sum_i \av{|{\bi u}_i|^2}/N$ 
is  a half of  the  plateau value $M_{\rm p}$ of the 
mean square displacement  at low $T$ (see Eq.(A5) below).  
  From the appendix,  
$\av{|{\bi u}|^2} $ increases with increasing  the system length $L$ 
logarithmically in 2D as \cite{Janco}    
\be
\av{|{\bi u}|^2} =  M_{\rm p}/2\cong 
C_0+ C_1 \ln (L/\sigma_1), 
\en 
where $C_0$ and $C_1$ are functions of $T$ and are independent of $L$.  
In terms of the  shear modulus 
 $\mu$ and and the bulk modulus $K$,  the coefficient $C_1$ is expressed as 
\be 
C_1= 
\frac{k_BT}{2\pi} \bigg(\frac{1}{\mu}+ 
\frac{1}{K+\mu}\bigg).   
\en  
 In our 2D  system, 
 we obtained  $\mu\cong 18$   and $K \cong 67.5$ 
 at $T=0.56$ from the inital linear growth  of 
stress-strain relation and the density-pressure 
relation (not shown in this paper) \cite{sound}. 
Then we find $C_1\cong 0.0060$ at $T=0.56$.  
If  the  total particle number is increased from  
$4000$ to  $64000$, Eq.(4.6) yields   the incremental increase   
$C_1\ln 4\cong 0.0083$ in  $\av{|{\bi u}|^2} $. 
In fair agreement with  this estimate,  we  numerically 
calculated    $\av{|{\bi u}|^2} $ 
to be 0.02266 for $N=4000$ and 
0.03305 for $N=64000$, where  
${\bar {\bi r}}_i$  was equated with 
the time-average of ${\bi r}_i(t)$ in a time interval of width 200.

With Eq.(4.6), 
we need to examine how the particles remain within or go outside  their  
initial circle.  In Fig.5, we  display   $\Delta r_i(t_0,t_1)$ 
 in Eq.(4.5)  for five particles in  time range  $t=t_1-t_0<18000$ 
 with $N=4000$. 
These particles   are separated from one another 
with  distances longer than 10.  We recognize that 
these displacements  undergo  rapid thermal fluctuations 
with magnitudes  nearly equal to  the overlap 
length   $0.3$.  In the early stage ($t <\tau_{bp}$), 
most of ${\cal {F}}_i$  
 fluctuate  between 1 and 0. In this example, 
two  of them  escape  from their  initial circle, 
where one has a broken bond   at 
 $t \sim 8000$  and the other has 
two broken bonds at $t\sim 10000$ (see Fig.6). 
Each jump itself occurs  on a short  time of order  10.  
In (a) of Fig.6 gives   the trajectory 
of the particle escaping from 
a  cage at $t\sim 10^4$ in Fig.5. 
In (b)  of Fig.6, 
this  escape   take place  as a  string-like  motion  
involving  several particles as in 3D \cite{Kob,Sc}.  See (a) 
 of Fig.13 below for other examples. 
In   (b) of Fig.6, we can see that 
 most of the particles involved  
 have only one broken bond, 
which is particularly the case for 
isolated configuration changes. 
Therefore, these motions may also  be treated   as small slips in 2D.


Removing the rapid temporal fluctuations, 
we calculated   the smoothed displacement lengths,
\be 
\Delta {\bar r}_i(t)= 
|{\bar{\bi r}}_i(t+t_0)-{\bar{\bi r}}_i(t_0)|, 
\en 
for the  time-smoothed positions,  
\be 
{\bar{\bi r}}_i(t)= \frac{1}{t_{\rm sm}} 
\int_0^{ t_{\rm sm} }dt' 
{{\bi r}}_i(t+t').  
\en 
In  the  inset of Fig.5, the smoothing time $t_{\rm sm}$ is   500.  
Even  on this  timescale, the particles move considerably, 
even   across their initial circle.  
 Note that  this  $t_{\rm sm}$ is much longer than  the traversal time 
of the transverse sounds across the system  
 $t_a= L/c_\perp \sim 17$ for $N=4000$ \cite{sound}. 
These complex  
  fluctuations  should originate  from  superposition   
of weakly  coupled,   low-frequency  vibration  modes  
\cite{La,Angela,Ganter,Shintani,Reichman,Ruocco,Barrat,Liu,Bonn,Br,Sc}. 

\begin{figure}[htb]
\includegraphics[width=0.9\linewidth, bb = 0 0 176 176]{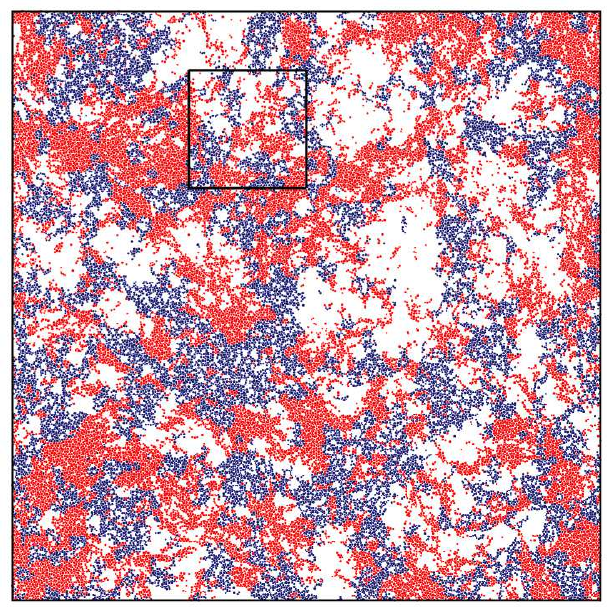}
\caption{Late-stage snapshot  
of particles with  ${\cal F}_i=0$ in a large system of 
 $N=64000$ and $L=281$, 
where  $t=10^4$ and $T=0.56$. Depicted 
particles  are classified into those with 
 ${\cal B}_i=0$ (red) 
and those with   ${\cal B}_i\ge 1$ (blue).  
  Numbers 
of the former (red) and  the latter (blue)  
are $(22034,18122)$, respectively. 
The former  arise from the  low-frequency 
vibration modes. Subsequent time-evolution in the upper box 
region will be given in the right of Fig.14. 
}
\end{figure}  
 
In Fig.7, 
we show the particles  with 
 ${\cal F}_i=0$ at very early times $t= 200$ and 400 for  $T=0.56$ 
and  $N=4000$, where their  initial  circles contain   no  particle.  
The fraction of the particles with   ${\cal F}_i=0$, 
written as  $\phi_4(t)$,  is soon about 0.2 for $t\gs 10$.  
 That is, a considerable amount 
of the  non-$\bi B$ particles 
with  ${\cal F}_i=0$ already appear from  very early 
times.  However, some of  the  $\bi B$ particles depicted 
  at $t=200$ are changed  to  
  those  with  ${\cal F}_i=1$ at $t=400$ (see 
the sentences at the end of Sec.IVA).

In Fig.8, 
we show  the particles  with 
 ${\cal F}_i=0$ at $t=10^4 \sim \tau_\alpha$ 
for   $T=0.56$ in a much larger system of $N=64000$.  
 Some heterogeneities have sizes of order 50. 
Among the displayed particles,   
  the $\bi B$ and non-$\bi B$  
particles  amount to  22034 and 18122, respectively. 
The patterns of the latter are more extended than those 
of the former. 
Moreover, their timescales  are  distinctly separated (see  Fig.14 below).
 These  indicate    
the presence of  thermally  excited  large-scale 
vibration modes. 

\subsection{Four-point correlations}
\begin{figure}[t]
\includegraphics[width=0.9\linewidth, bb=0 0 433 252]{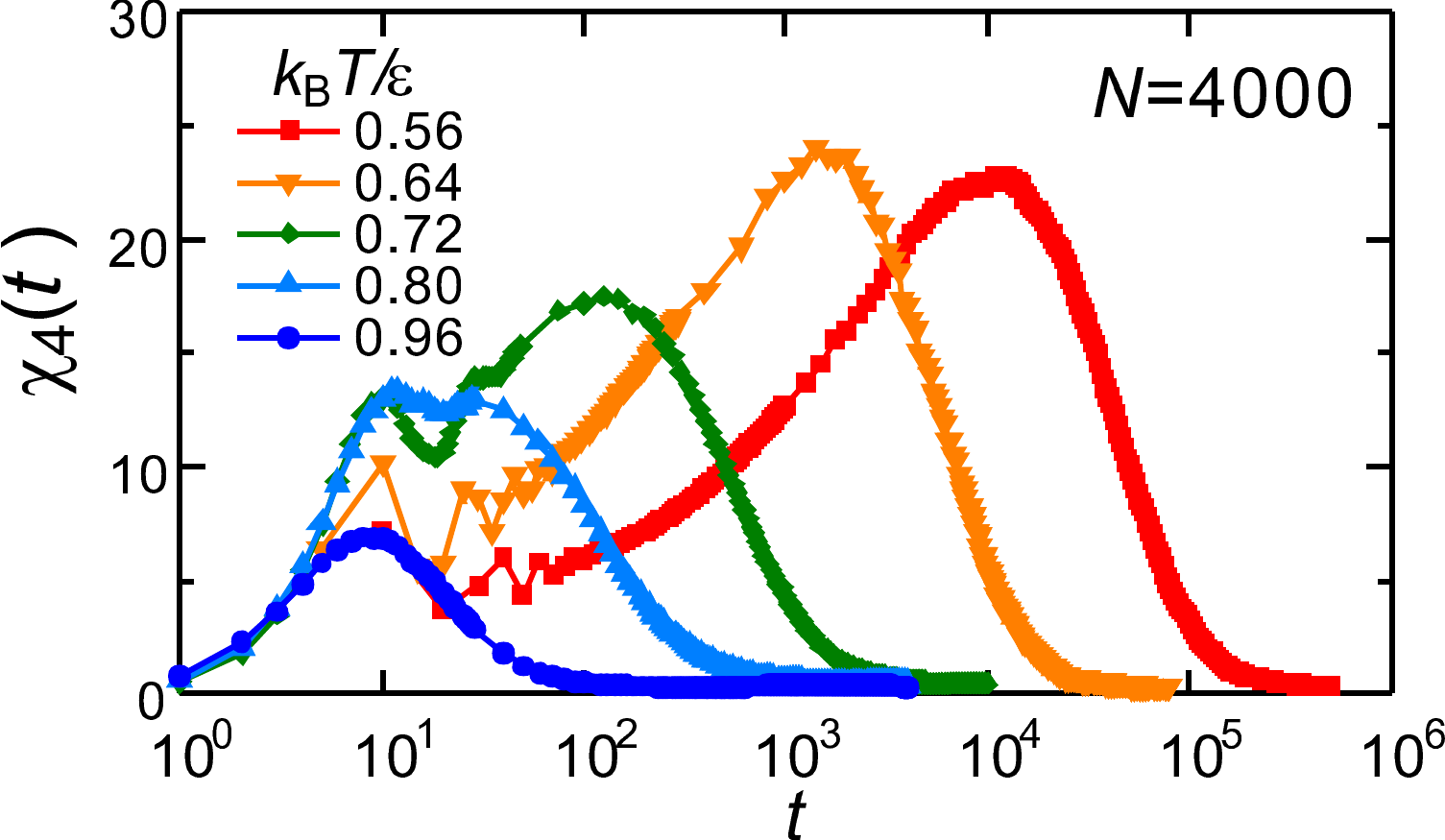}
\caption{Susceptibility  $\chi_4(t)$ in Eq.(4.14) 
vs $ t$  for $N=4000$. }
\end{figure}
\begin{figure}[t]
\includegraphics[width=0.9\linewidth, bb=0 0 434 494]{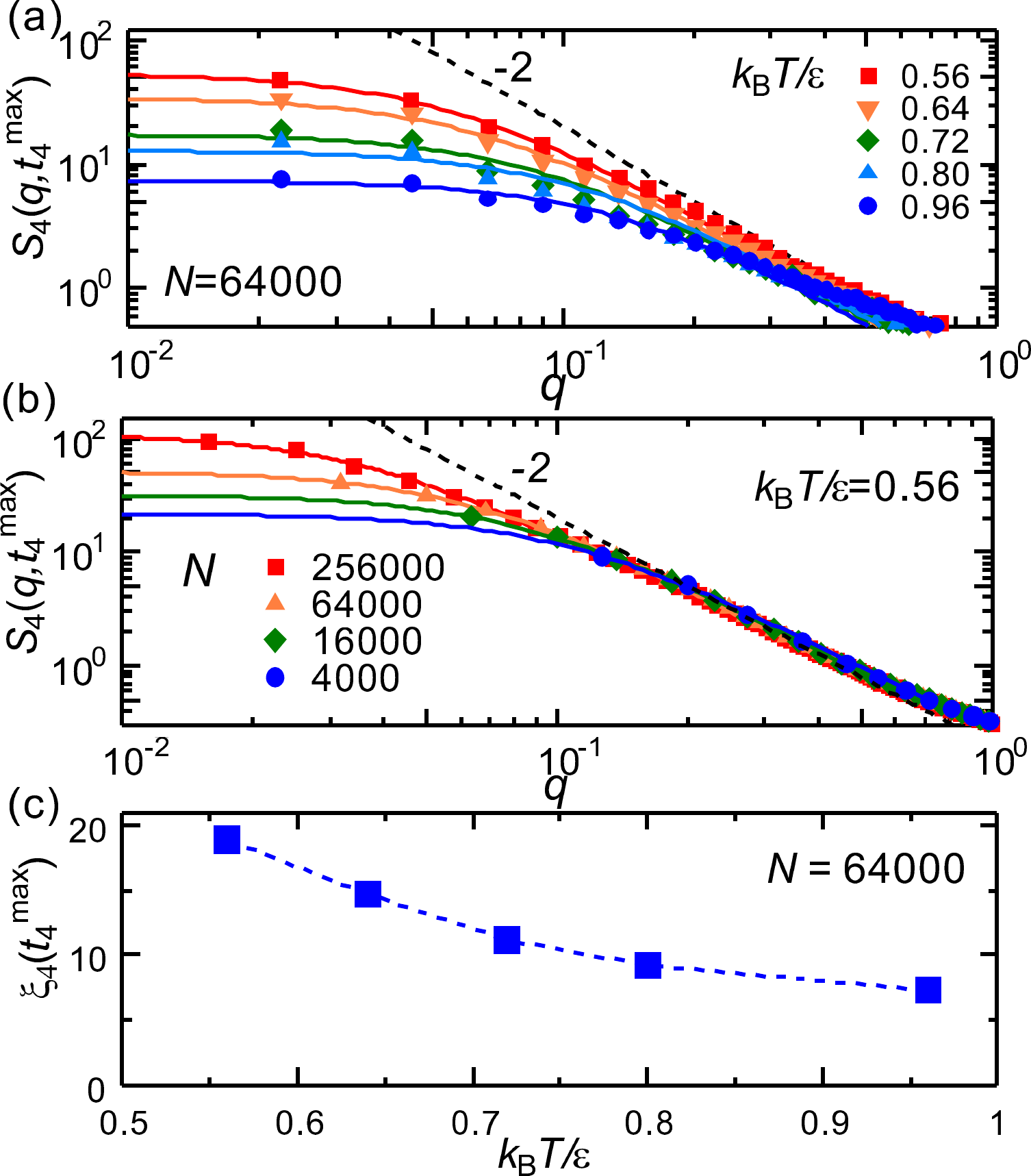}
\caption{(a)  Four-point structure factor 
$S_4(q,t_4^{\rm max})$ vs $q$ in Eq.(4.13)    
 for various $T$, where $N=64000$. (b) $S_4(q,t_4^{\rm max})$ vs $q$ 
for various $N$ at $T=0.56$. (c) $\xi_4 (t_4^{\rm max})$ for $N=64000$.  
Marked  system-size dependence appears  at  small $q$  
due to the low-frequency vibration modes. }
\end{figure}

 The  average  of 
$\hat{\cal{Q}}({\bi r},t_0,t_1)$ in Eq.(4.1) 
 is written as 
\bea 
q_4(t)&=&
 \av{\hat{\cal {Q}}({\bi r},t_0,t_1)} \nonumber\\
&=& \frac{1}{V}\sum_{ij}
w(|{\bi r}_i(t_0)-{\bi r}_j(t_1)|) ,
\ena 
which is a function of  $t=t_1-t_0$. In our simulation, 
${\cal F}_i= 1$ or 0, so $q_4(t)$ is related to   the fraction 
 $\phi_4(t)$  of the 
particles with ${\cal F}_i=0$  as  
\be 
q_4(t)= n[1- \phi_4(t)].
\en 
Here,  $q_4(0) = n$ 
and $q_4(\infty)= v_0 n^2$, 
where $v_0$  is the  area or volume of the overlap region 
($v_0=\pi (A_4\sigma_1)^2$ in 2D and 
$v_0= 4\pi (A_4\sigma_1)^3/3$ in 3D).

As   in Eq.(3.19) for  $G_b({ r},t)$, 
 the  four-point  space-time correlation function is given  by  
\bea 
G_4({ r},t)&=&
 \av{\hat{\cal {Q}}({\bi r}+{\bi r}',t_0,t_1) 
\hat{\cal{Q}}({\bi r}',t_0,t_1)} 
\nonumber\\
&&\hspace{-2.2cm} 
=\frac{1}{V}\Av{
\sum_{ik}  
{\cal F}_i (t_0,t_1) {\cal F}_k (t_0,t_1)  
\delta ({\bi r}- {\bi r}_{ik}(t_0)) },
\ena 
where ${\bi r}_{ik}(t_0)
={\bi r}_i(t_0)-{\bi r}_k(t_0)$. 
The   four-point structure factor is defined by  
\bea
S_4(q,t)&=&  \frac{1}{V}  \av{|{\hat{\cal Q}}_{\bi q}(t_0,t_1)|^2}\nonumber\\
&&\hspace{-1cm}=
 \int d{\bi r}[G_4({r},t)-q_4(t)^2] 
e^{i{\bi q}\cdot{\bi r}} , 
\ena
where ${\hat{\cal Q}}_{\bi q}(t_0,t_1)= 
\sum_{j}{\cal F}_j (t_0,t_1) 
 \exp[i{\bi q}\cdot{\bi r}_j(t_0)]$ is the 
Fourier component of ${\hat{\cal Q}}({\bi r}, t_0,t_1)$. 
We define the four-point susceptibility   
$\chi_4(t)$ by 
\be 
\chi_4(t)=\frac{1}{V} \Av{
\sum_{ik} \delta{\cal F}_i (t_0,t_1) \delta{\cal F}_k (t_0,t_1) }. 
\en 
in terms of the deviation 
 $\delta {\cal F}_i (t_0,t_1)= 
{\cal F}_i (t_0,t_1)- q_{4}(t)/n$. 
In these correlation functions, the four particle 
positions ${\bi r}_i(t_0)$, ${\bi r}_k(t_0)$, 
${\bi r}_j(t_1)$, and ${\bi r}_\ell(t_1)$ 
are involved. However, as discussed around Eq.(4.4), 
  the self parts with $i=j$ and $k=\ell$ 
dominate over the distinct parts with  
 $i\neq j$ and $k\neq \ell$ \cite{Lacevic}. 

 Berthier 
{\rm et al.}\cite{b1,b2} showed that the four-point susceptibility  
$\chi_4(t)$ depends on the ensemble ($NVE$ or $NVT$) 
 and the dynamics (Newtonian or Brownian). 
Our $\chi_4(t_4^{\rm max})$   in the $NVE$ ensemble  
 is  roughly $60\%$  of the long wavelength limit 
 of  the four-point structure factor $\chi^0_4(t_4^{\rm max})= 
\lim_{q\to 0}S_4(q,t_4^{\rm max}) $, which is consistent  with 
  the previous calculations   
\cite{b1,b2,Chandler,Szamel,Sas}. 
However, in our calculation, 
  the long wavelength limit 
 of  the bond-breakage structure factor $\chi^0_b(t_b^{\rm max})= 
\lim_{q\to 0}S_b(q,t_b^{\rm max}) $ 
cannot be determined reliably because of the 
very long $t_b^{\rm max}$ at low $T$ 
(see Fig.4) and  our maximum  
bond-breakage susceptibility 
 $\chi_b(t_b^{\rm max})$ in Fig.3 
apparently exceeds $\chi^0_b(t_b^{\rm max})$ 
by a few ten $\%$.  Thus,  
we cannot draw a definite 
conclusion on the relation between 
 $\chi_b(t_b^{\rm max})$  and $\chi^0_b(t_b^{\rm max})$.

 In (a)  of Fig.9, we     give     
 $\chi_4(t)$ in Eq.(4.14)  as  a function of 
 $t$ for  various $T$, which is calculated in 
  the $NVE$ ensemble with  $N=4000$. It  
is  maximized at  $t= t_4^{\rm max}$.  
Here, due to  the transverse sound  propagation, 
 a smaller acoustic peak emerges 
with lowering $T$  at $t= t_a/2 \sim 8.6$ \cite{sound}, 
whose existence  has not been reported in the previous papers. 
It becomes  more evident   at lower $T$,  
where the  acoustic damping is  weaker.
We also calculated $\chi_4(t) $ for 
other $N$.  For  $N=64000$,   
the acoustic  peak was 
at $t\sim 34$ and its  height 
 even exceeded  the first peak height   
  for low $T$. For $N=1000$,  
there was no acoustic   peak. 
See the item (3) in the summary  for more 
discussions.

La\v{ c}evi\'{c} {\it et al.}\cite{Lacevic} 
determined   the overlap length  $a\sigma_2$ 
 to maximize the 
peak height of the four-point susceptibility     
$\chi_4(t_4^{\rm max})$    
as  a function of the parameter  $a$ at  $T=0.59$ in 3D.  
They then obtained 
 $a=0.3$ and used it also  
at other low  $T$.  
Following their method, we also maximized 
$\chi_4(t_4^{\rm max})$ as a function of the overlap length 
to obtain $A_4 =0.3$ in Eq.(4.3) for  $T=0.64$ and $N=4000$. 
Then $t_4^{\rm max} \sim  10^4$ at $T=0.56$ 
both for $N=4000$ and 64000.

In (a)  of  Fig.10, we plot $S_4(q,t_4^{\rm max})$ vs $q$ 
  for   various $T$ with $N=64000$. In its calculation, we took the average 
over the initial time $t_0$ in  the wide range $[0, 10^6]$. 
As in the case of $S_b(q,t)$  in Eq.(3.20),  
we may  fit  $S_4(q,t)$ to  the 
 Ornstein-Zernike form as \cite{Lacevic}  
 \be 
 S_4(q,t) =\chi_4^0(t)  /[1+q^2\xi_4(t)^2],   
 \en 
 where  $\chi_4^0(t)= \lim_{q \to 0}S_4(q,t)$ 
is the long wavelength limit of $S_4(q,t)$  and 
$\xi_4(t)$ is   the four-point 
correlation length.  Furthermore, 
in (b), we show  $S_4(q,t_4^{\rm max})$ at $T=0.56$  
 for various $N$ up to $N=256000$  to demonstrate 
its  significant   system-size dependence at small $q$.  
In (c),  we display   $\xi_4(t_4^{\rm max})$ vs $T$ 
for $N=64000$ as an example, which increases 
with lowering $T$. 
We recognize that the ratio 
$\xi_4(t_4^{\rm max})/\xi_b(t_b^{\rm max})$  
  exceeds unity  and 
 increases with increasing $N$. 
For example, it is about 3  
for  $T=0.56$ and $N=64000$. Thus, 
on our 2D simulation,  the effect of   the  low-frequency vibration 
modes on  the four-point correlations 
 becomes stronger for larger  $N$.

\begin{figure}[t]
\includegraphics[width=0.9\linewidth, bb = 0 0 462 301]{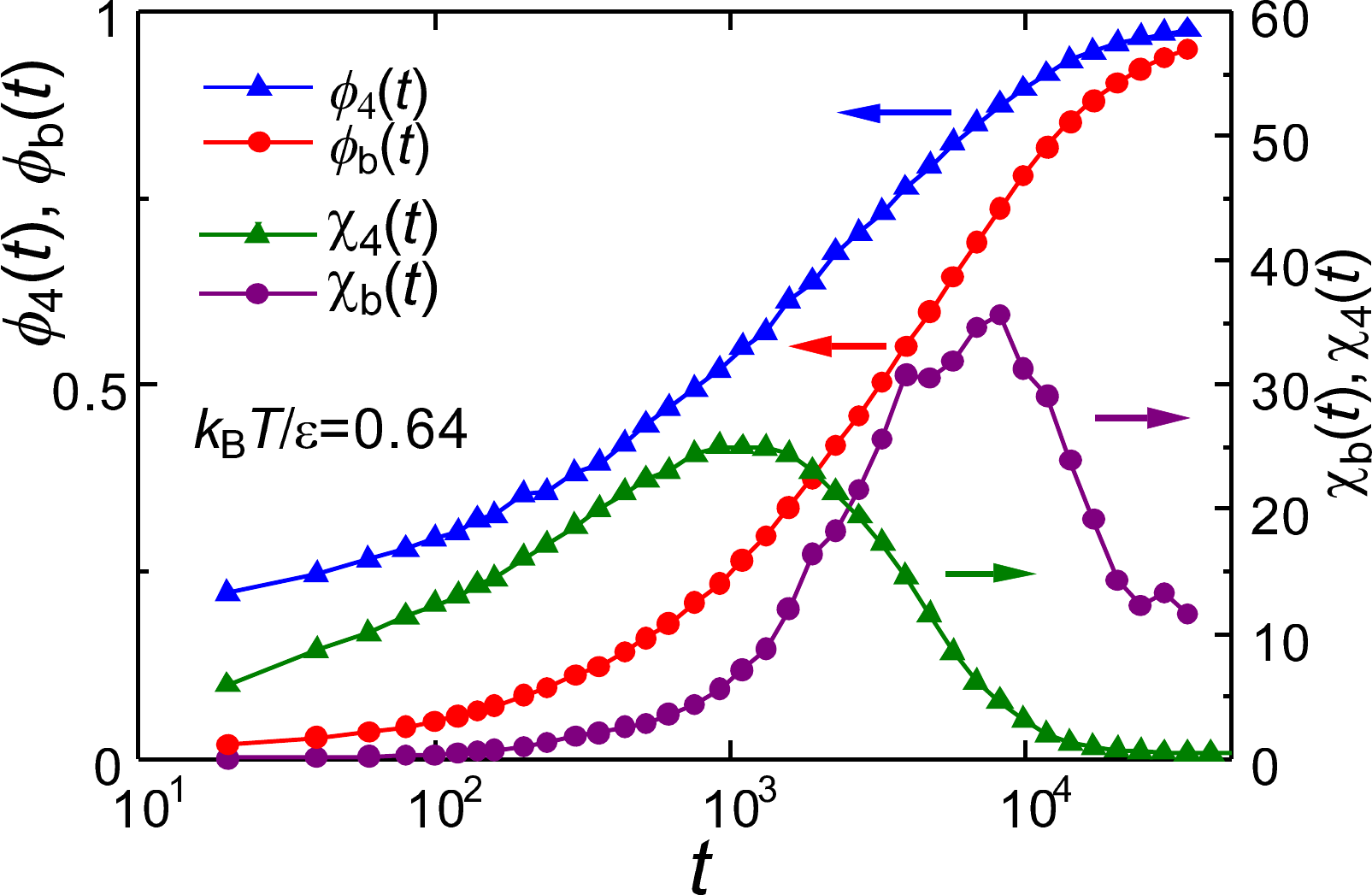}
\caption{ Fraction $\phi_b(t)$ and susceptibility  $\chi_b(t)$ 
for bond breakage and those 
$\phi_4(t)$ and $\chi_4(t)$ 
for four-point correlations for  $T=0.64$ and  $N=4000$.  
Here,  $\chi_b(t)$ and  $\chi_4(t)$ 
are maximized  for $\phi_b(t) \sim 1/2$ and  $\phi_4(t) \sim 1/2$, 
respectively.  
}
\end{figure}

\begin{figure}[t]
\includegraphics[width=0.8\linewidth, bb = 0 0 460 488]{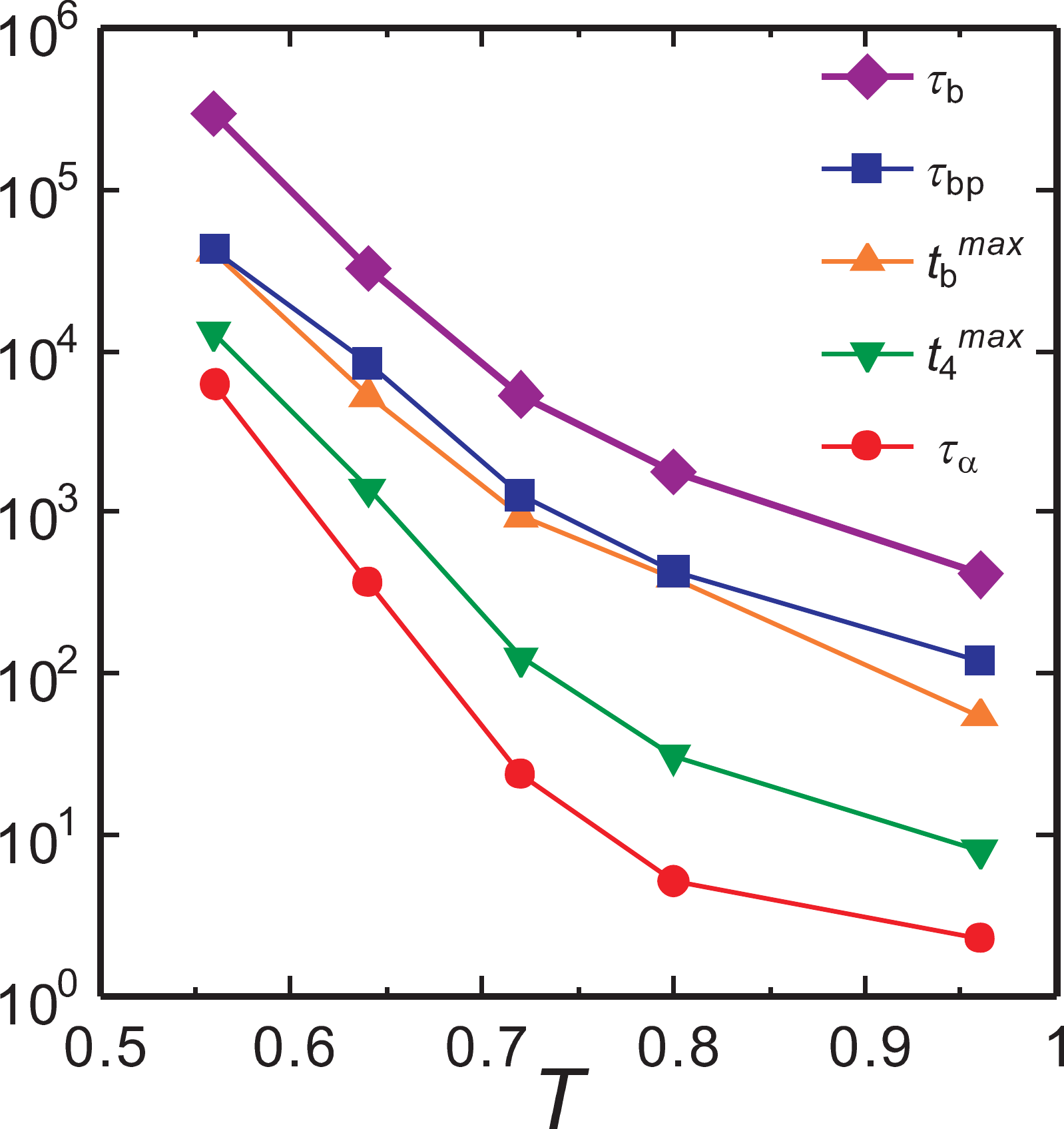}
\caption{Relaxation times as functions of  $T$ for 
the soft-core potential  for $N=4000$  in 2D. From above, they  are 
$\tau_b$ in Eq.(3.4),  $\tau_{bp}$ in Eq.(3.16), 
 $t_b^{\rm max}$ from the maximum of $\chi_b(t)$, 
 $t_4^{\rm max}$  from the maximum of $\chi_4(t)$, and  
$\tau_\alpha$ in Eq.(3.6).  
}
\end{figure}

\begin{figure*}
\includegraphics[width=1\linewidth, bb = 0 0 522 204]{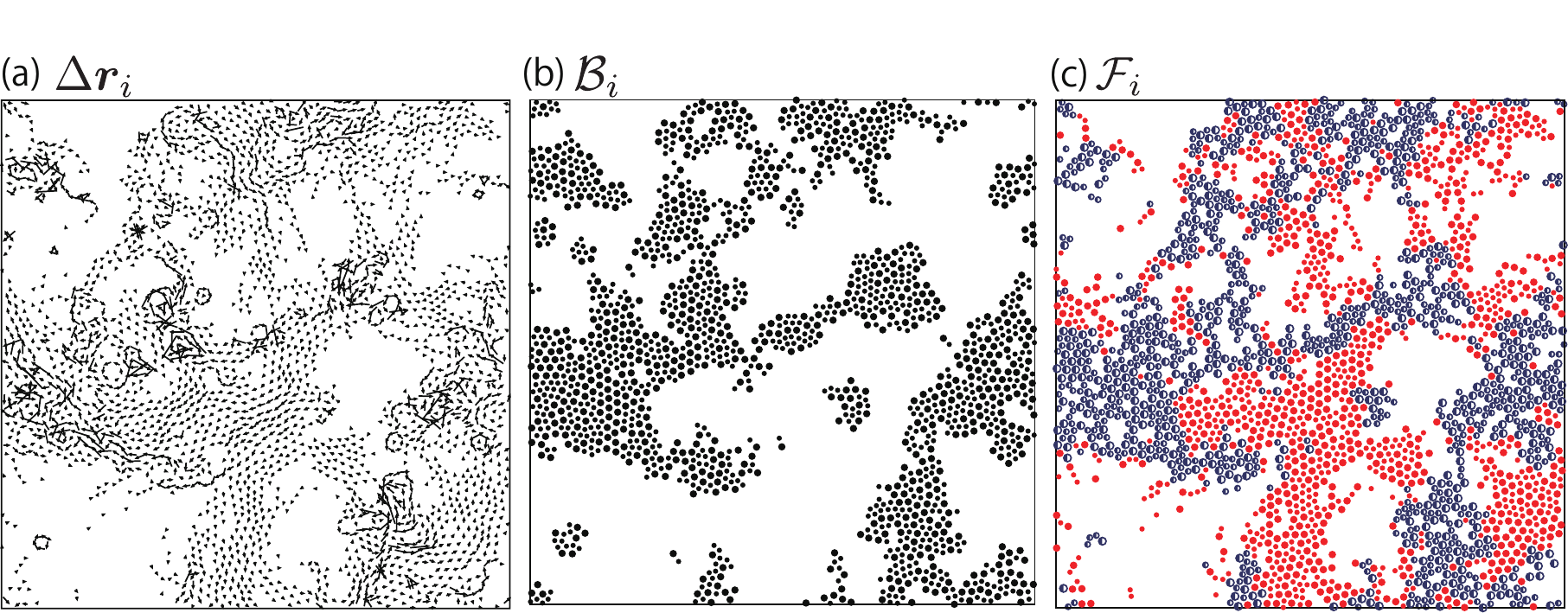}
\caption{
Snapshots at  $t=t_1-t_0=10^4$ for $T=0.56$ and $N=4000$ in 2D.  
System length is  $L=70.2$.  (a)    
Displacements $\Delta{\bi r}_i(t_0, t_1)$  
with $|\Delta{\bi r}_i|>0.3$,      (b)  $\bi B$ particles,  
and  (c) those with ${\cal F}_i=0$ 
 classified into  $\bi B$ 
 particles  (blue) and 
 non-$\bi B$ particles  (red). 
Non-${\bi B}$  
particles with ${\cal F}_i=0$ are 
produced by collective motions, while most  $\bi B$ particles 
 participate in    string-like  motions 
and  satisfy  ${\cal F}_i=0$. }
\end{figure*}

\begin{figure*}[htb]
\includegraphics[width=0.9\linewidth, bb = 0 0 732 404]{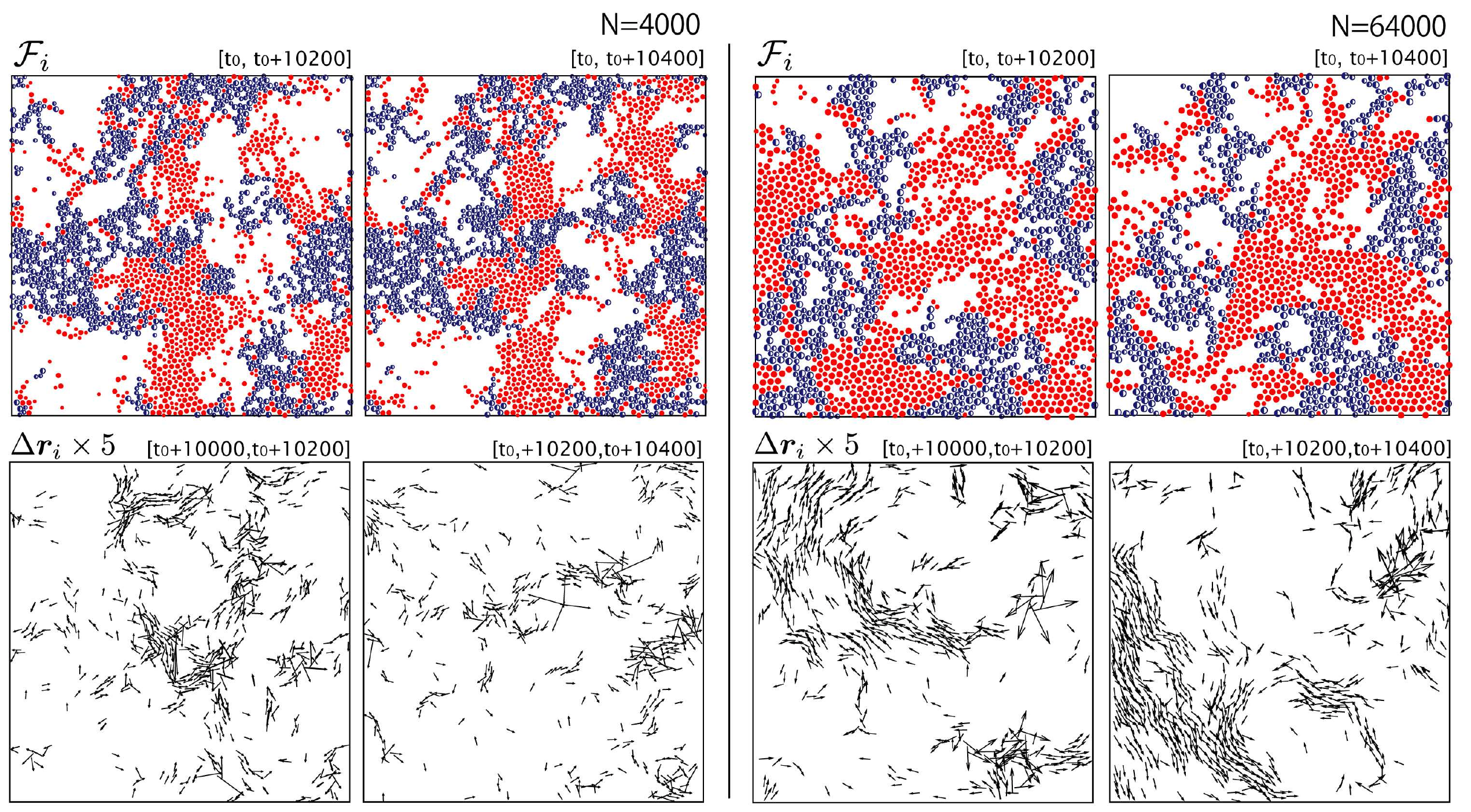}
\caption{Snapshots  at 
  $t= 10000+ 200 k$ ($k=1,2$)  at $T=0.56$ for 
$N =4000$ (left) and for $ N =64000$ (right).  
Top: Particles  with ${\cal F}_i=0$ 
 classified   into   non-$\bi B$ particles   (red) and $\bi B$ 
 particles (blue),  
where  the former   significantly change 
but the latter   little   change in this short time interval. 
The change of the former is larger  for the larger $N$ in the right.   
See the preceding  snapshots   in (c) of Fig.13 for $N=4000$ 
and  in the black box of Fig.8   for $N=64000$. 
 Bottom: Displacements $\Delta{\bi r}_i= \Delta{\bi r}_i(t_0+t', 
t_0+t'+200)$  
multiplied by 5 with $|\Delta{\bi r}_i|>0.3$ in  short time intervals  of 200, 
where $ t'=10000$ or 10200.  
}
\end{figure*}

\setcounter{equation}{0}
\section{Comparison of the two theoretical  schemes }

\subsection{Timescales }

In Fig.11,  we display   $\phi_b(t)$, $\phi_4(t)$,  $\chi_b(t)$, and  
$\chi_4(t)$  at $T=0.64$ for $N=4000$ in  the two theoreticaal schemes. 
Here, $\phi_b(t)$ is the fraction of the particles with 
${\cal B}_i>0$ and  $\phi_4(t)$ is that  with 
${\cal F}_i =0$.   
We can see that  $\phi_b \sim 1/2$ at $t= t_b^{\rm max}$  
 and $\phi_4 \sim 1/2$ at $t= t_4^{\rm max}$. If $\phi_b(t)$ (or $\phi_4(t)$) 
is close to 0 or  1, $S_b(q,t)$ (or $S_b(q,t)$) becomes 
very small.

So far we have introduced 
the bond-breakage time $\tau_b$  
 in Eq.(3.4), the relaxation time 
$\tau_\alpha$ from $F_s(q,t)$ in Eq.(3.6), 
the bond-preserving time $\tau_{bp}$ in Eq.(3.16), 
the maximization time  $t_b^{\rm max}$ of   $\chi_b(t)$ 
in (a)  of Fig.3, 
 and the maximization 
  time $t_4^{\rm max}$  of   $\chi_4(t)$ in Fig.9.
In Fig.12, these times are  in the following order, 
\be 
 \tau_\alpha <t_4^{\rm max}< t_b^{\rm max}\sim \tau_{bp}<
\tau_b , 
\en 
 with $35\le 
\tau_b/\tau_\alpha\le 10^2$ in the range $0.56\le  T\le 0.96$ for 
$N=4000$.  In our 2D simulation, 
$\tau_\alpha$ and $t_4^{\rm max}$ 
exhibit  strong system-size dependence 
due to the low-frequency vibration modes.  
In fact, 
we  numerically obtained  
  $\tau_\alpha=8400$   for $N=4000$ and $\tau_\alpha=2140$ 
for $N=64000$ (see the  appendix). 
It is worth noting that 
significant system-size dependence of the plateau  
behavior  of $F_s(q,t)$ (and  $\tau_\alpha$ from Eq.(3.6))  
has been reported  in 2D and 3D simulations 
\cite{finite,Sas,Kim}. Furthermore,  
Karmakar {\it et al.} \cite{Sas}  examined  
  system size-dependence of $\xi_4$ and $S_4(q,t)$ 
up to $N=351232$  in 3D.

In addition, 
we comment on the  stress-time-correlation function, 
which  considerably decreases    in the  early  stage 
due to the thermal motions  as well  as 
$F_s(q,t)$.  As a result, 
its relaxation time is of order   $\tau_\alpha$  
\cite{Furukawa}, while the nonlinear flow 
behavior is characterized by $\tau_b$ \cite{yo1}.

\subsection{Time-evolution  on long and short timescales}

In (a)  of Fig.13,  the arrows represent   
 relatively large displacements $\Delta{\bi r}_i= 
{\bi r}_i (t_1)-{\bi r}_i(t_0)$  
with $|\Delta{\bi r}_i|>0.3$ \cite{Hi,Dol}. 
We can see both large-amplitude string-like motions 
and smaller-amplitude collective motions.  
In (b),   all the   $\bi B$ particles are displayed. 
In (c),  those  with 
 ${\cal F}_i=0$ are divided into 
 $\bi B$    and  non-$\bi B$ particles, where the former       
exhibit patterns closely resembling 
those of the ${\bi B}$ particles in (b). 
The number of the $\bi B$ particles with 
${\cal F}_i=0$ is  about   $70\%$  of that 
 of the total $\bi B$ particles in (b).

From (a) and (c) of  Fig.13,  
the  non-${\bi B}$ particles 
with ${\cal F}_i=0$    
mostly  arise from the collective motions on large scales, 
 as has been  the case at short times in   Fig.7. 
Their selection is 
very sensitive to the overlap length, $a\sigma_2$ in the original 
work\cite{Lacevic} and $A_4\sigma_1$ 
in this paper, while  the $\bi B$  
 particles (even with  ${\cal F}_i=0$)    
 are relatively  insensitive to it. 
 In  the four-point theory \cite{Lacevic},  
the overlap length was chosen to maximize 
$\chi_4^{\rm max} =\chi_4 (t_4^{\rm max})$, as discussed above Eq.(4.15).   
Roughly speaking, their method  is  to maximize  the contribution from  the 
 thermal  collective motions  
of non-${\bi B}$  particles to    $\chi_4(t)$ (see Fig.11).

\begin{table}
\caption{Particle numbers with (a)  ${\cal B}_i>0$, 
(b) ${\cal B}_i>0$ and ${\cal F}_i=0$,  
 (c) ${\cal B}_i={\cal F}_i=0$, and 
 (d) $\Delta r_i>0.3$  in Fig.14 
at $t=10200$ and 10400 for 
$N=4000$ and 64000 in 2D. 
Those of the particles  common  in 
 these two shots 
are also given. Subscript $i$ is 
omitted from $\Delta r_i$, $ {\cal B}_i$, 
and ${\cal F}_i$.}
\begin{tabular}{|c|c|c|c|c|} 
\hline
$t$ ($N$)  & 
${\cal B}>0~~$ & ${\cal B}>0$,${\cal F}=0$  & 
${\cal B}={\cal F}=0$ & $\Delta r>0.3$\\
\hline
$10200$(4000)  & 1629 & 1104 & 1091&2451 \\
\hline
$10400$(4000)  & 1667 &1180 & 1221& 2643\\
\hline
common(4000)  & 1576 & 925 & 823 & 1923   \\
\hline\hline 
$10200$(64000)  & 26188 & 18336 & 20884&43544  \\
\hline 
$10400$(64000)  & 26413& 18678 & 21785 & 44693\\
\hline 
common(64000)  &25192 & 14149 & 14349&    31790  \\
\hline
\end{tabular}
\end{table}

\begin{table}
\caption{Numbers of non-$\bi B$ and $\bi B$ particles for three 
categories   in Fig.14 at $t=10200$ for 
$N=4000$ in 2D.   }
\begin{tabular}{|c|c|c|c|c|} 
\hline
\hspace{-3mm} 
& 
  ${\Delta r}>0.3$  & 
${\Delta r}>0.3$& ${\Delta r}<0.3$ &  \\
& 
   $ {\cal F}=0$ & 
$ {\cal F}=1$& ${\cal F}=1$& total \\
\hline
${\cal B}=0$ & 1091 & 7 &   1273 & 2371 \\
\hline
${\cal B}>0$&1104  &  249&   276& 1629   \\
\hline
total & 2195 &  256 &1549 & 4000 \\
\hline
\end{tabular}
\end{table}

We next examine time-evolution 
 at two consecutive  times  $t=10200$ and 10400 
for $N=4000$ (left) and 64000 (right) at   $T=0.56$. 
The four upper panels of Fig.14  display   the particles 
with  ${\cal F}_i(t_0,t_0+t)=0$  
grouped  into  $\bi B$ 
 and  non-$\bi B$ particles.  
 These  snapshots are  subsequent  to 
 that in (c) of Fig.13  for $N=4000$ and 
that in the box region in Fig.8 for  $N=64000$  
  in  the same runs.  The four lower  panels of Fig.14 give 
the corresponding displacements $
\Delta{\bi r}_i(t_0+t', t_0+t'+200)$ 
with  $t'=10000$ or 10200, which  exceed 
 0.3  in this  short time intervals of 200.   
  The differences  
between  the consecutive patterns 
are evidently larger for $N=64000$ than for  $N=4000$. 
We can see that this system-size dependence 
originates  from   
the  large-scale vibrational  motions.  

In Table I, 
we give the numbers of the  particles 
with (a) ${\cal B}_i>0$, 
(b) ${\cal B}_i>0$ and ${\cal F}_i=0$, 
(c) ${\cal B}_i={\cal F}_i=0$, and (d) $\Delta r_i>0.3$ 
 in the four snapshots in 
Fig.14.  We also give those of the particles  commonly depicted in 
 the consecutive snapshots. 
 In Table II,  the numbers  of  the $\bi B$ and non-$\bi B$ particles 
 are given  at $t=10200$ for  (a) ${\Delta r}_i>0.3$ 
 and ${\cal F}_i=0$, 
(b) ${\Delta r}_i>0.3$ 
 and ${\cal F}_i=1$,
and (c) ${\Delta r}_i<0.3$ 
 and ${\cal F}_i=1$. There is no particle with 
${\Delta r}_i<0.3$  and ${\cal F}_i=0$.   
We recognize the following. (i) About $20-30\%$ of the particles 
with  ${\cal F}_i=0$  change into 
those with  ${\cal F}_i=1$ and vice versa in  a short time of 200. 
(ii) About $30\%$ of the ${\bi B}$ particles 
satisfy ${\cal F}_i=1$ at each time  because 
of  the presence of another particle  $j$ 
within their initial circles. As stated at the 
end of Sec.IVA, most of 
the particles with $\Delta r_i>0.3$ and ${\cal F}_i=1$  
are $\bi B$ particles having broken bonds (which  is $97\%$ 
in the example of Table II). 
(iii) About $5\%$ of the $\bi B$ particles 
become the non-$\bi B$ particles and  vice versa  
in a short time of $200$. 
(iv) About $85\%$  of the $\bi B$ particles 
move outside their initial circle 
to have $\Delta r_i>0.3$. The remaining $15\%$ $\bi B$ particles 
stay within their  initial circle having  
broken bonds  after   long-distance  motions  
of the neighboring   particles.

 Dauchot {\it et al.} \cite{Dauchot} performed an  experiment   
  on a 2D dense granular packing  
 under cyclic shear near the jamming transition. 
Their  snapshots of $\Delta {\bi r}_i$  and 
$1-{\cal F}_i$ ($\hat{q}_s^a$ in their notation) 
  resemble  those in  Fig.13.

\section{Three-dimensional results}

Also in  3D,  the four-point corelations arise from  
the bond-breakage motions and  the  thermal vibrational  motions. 
The former grow slowly with structural 
relaxations, while  the latter fluctuate 
relatively rapidly.

\begin{figure*}
\includegraphics[width=0.9\linewidth, bb = 0 0 557 199]{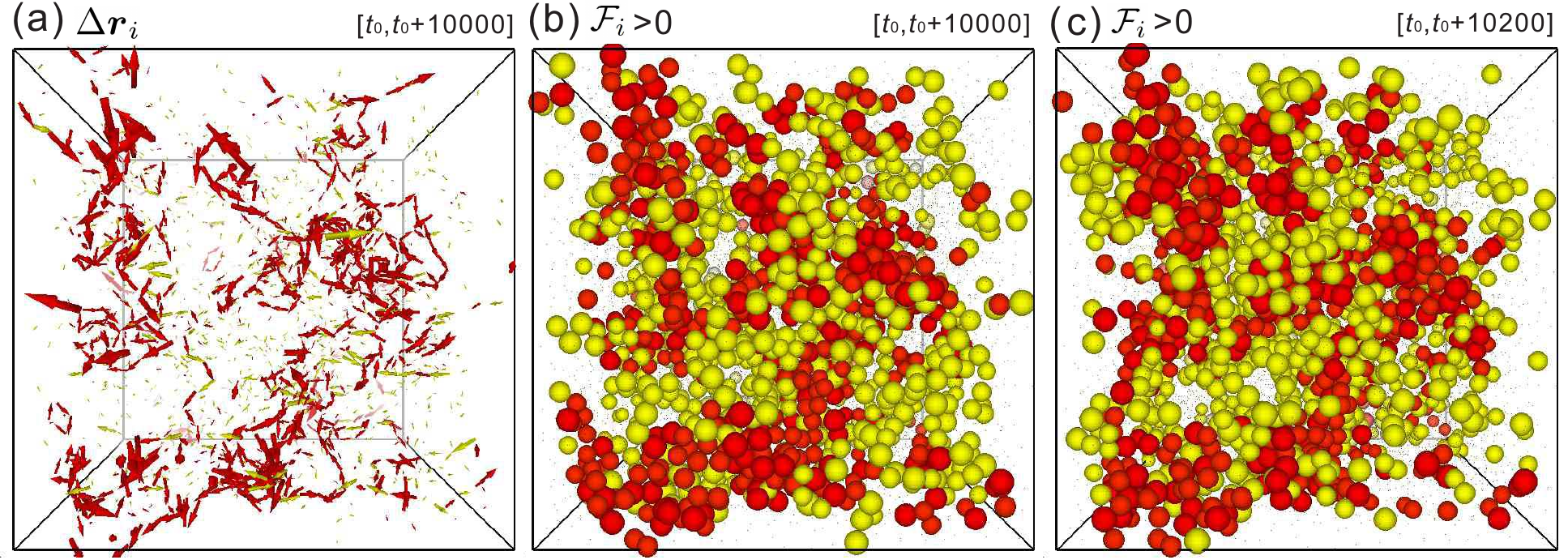}
\caption{
Snapshots  in 3D  for $T=0.24$ and  $N=10000$.    
(a) Arrows indicate 
displacements $\Delta{\bi r}_i(t_0, t_0+10000)$  
with $|\Delta{\bi r}_i|>0.3$, whose number is 1818. 
 Particles  with ${\cal F}_i=0$ 
 classified into  $\bi B$ 
 particles   (red) and  
non-$\bi B$ particles   (yellow)   
for time intervals $[t_0,t_0+10000]$  in (b) 
and $[t_0,t_0+10200]$ in (c). 
 Depicted non-${\bi B}$  
particles  are 
produced by the vibration  modes 
 and are fluctuatiing in time, while 
 $\bi B$ particles are not much changed between these two times.
}
\end{figure*}

\begin{figure}
\includegraphics[width=0.9\linewidth, bb = 0 0 406 664]{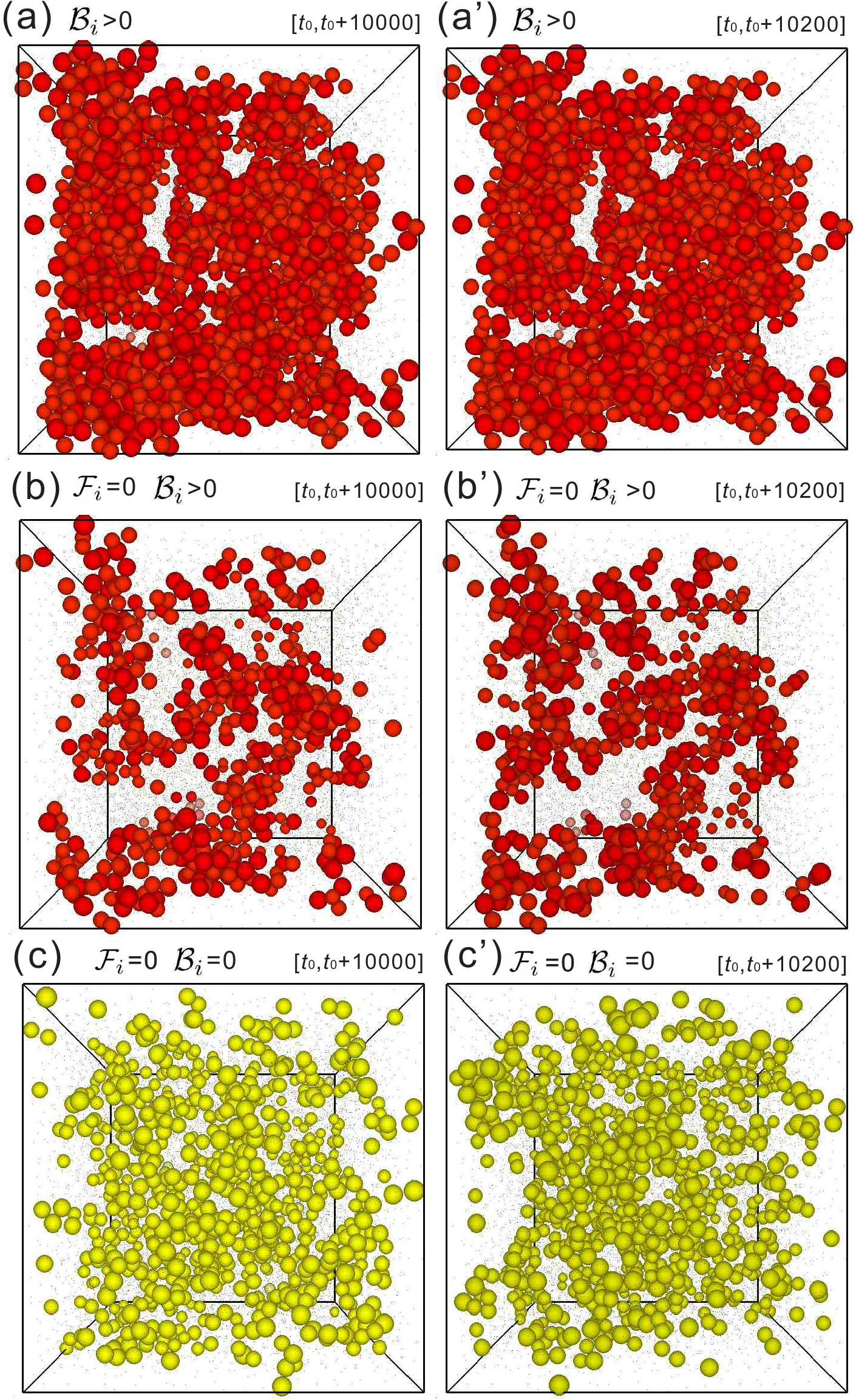}
\caption{ 
Snapshots using the data  in Fig.15 
for  time intervals $[t_0,t_0+10000]$ (left) 
and $[t_0,t_0+10200]$ (right). 
Displayed are particles with 
  ${\cal B}_i>0$ in  (a), those with 
${\cal B}_i>0$ and ${\cal F}_i=0$ in (b), 
and those with 
${\cal B}_i={\cal F}_i=0$ in (c).  
}
\end{figure}

In Fig.15, we give 
snapshots of the particles at $t=10000$ and 10200   
for  $T=0.24$ and  $N=10^4$, 
where  $\tau_\alpha \sim 10^5$. 
The system length is $L=23.2$. 
Here, the fraction of the $\bi B$ particles is $\phi_B(t)\sim 0.24$ 
and that  with ${\cal F}_i=0$  is 
  $\phi_4(t)\sim 0.15$. 
In (a),  we display    
the relatively large displacements $\Delta{\bi r}_i= 
{\bi r}_i (t_0+t)-{\bi r}_i(t_0)$ 
with $\Delta{ r}_i>0.3$. 
 String-like motions  
are conspicuous \cite{Kob,Sc,Lacevic,yo1}, around which 
 collective motions 
with  $|\Delta{\bi r}_i|>0.3$  tend to be  induced. 
We also display the particles with ${\cal  F}_i=0$ 
grouping them into 
  $\bi B$ particles  (in red)
and    non-$\bi B$ particles   (in yellow) 
in  time intervals  $[t_0,t_0+10000]$  in (b)   
and  $[t_0,t_0+10200]$ in (c).

In Fig.16, we  
display  the $\bi B$ particles  in (a) and (a'),   
 the $\bi B$ particles with ${\cal F}_i=0$  
in  (b) and (b'), and 
 the non-$\bi B$ particles with ${\cal F}_i=0$  
in  (c) and (c') 
for $t=10000$ in the left and 10200 in the right. 
We use   the same data as in   Fig.15.  
We can see that     
the  $\bi B$ particles little  change, 
but the non-$\bi B$ particles 
much change in a time interval of 200. 
The aggregates of  the ${\bi B}$ particles 
have grown from the strings in (a) in  Fig.15. 
The number of     
the  total $\bi B$ particles (in the top panels)  
is considerably larger than that of  the $\bi B$ particles with 
${\cal F}_i=0$ (in the midde panels),
 which are 2351 and 641, respectively, at $t=10000$. 
This is because the particles surrounding each string 
can have broken bonds without their long-disance motions. 

Table III  gives  the numbers of the  particles 
with (a) ${\cal B}_i>0$, 
(b) ${\cal B}_i>0$ and ${\cal F}_i=0$, 
(c) ${\cal B}_i={\cal F}_i=0$, and (d) $\Delta r_i>0.3$ 
 in the two  snapshots in 
Figs.15 and 16.  Also given in the last line are  
 those of the particles  commonly depicted in 
 the consecutive snapshots. On the other hand, 
 Table IV presents 
 the numbers  of  the $\bi B$ and non-$\bi B$ particles 
  at $t=10200$ for  (a) ${\Delta r}_i>0.3$ 
 and ${\cal F}_i=0$, 
(b) ${\Delta r}_i>0.3$ 
 and ${\cal F}_i=1$,
and (c) ${\Delta r}_i<0.3$ 
 and ${\cal F}_i=1$. Conspicuous features are as  follows.  
(i) About $50\%$  particles with $\Delta r>0.3$ 
change into those with   $\Delta r_i<0.3$ and vice versa 
in a short time interval of 200.  
Only  $30\%$ of the non-$\bi B$ particle 
are  common  in the two consecutive snapshots. 
(ii)  About $60\%$ of the $\bi B$ particles satisfy 
 $\Delta r_i<0.3$ and ${\cal F}_i=0$. (The  corresponding 
percentage is about $15\%$ in 2D in Table II.) 
This is because of the larger  
coordination number or the larger  bonded   particles 
around each particle in 3D.  That is, there are about 10 bonded 
particles around a particle 
which has participated in a string-like motion 
and moved over a molecular distance. 
 (iii)  The particles with 
 $\Delta r_i>0.3$   and ${\cal F}_i= 1$ 
amount  to $366$  
and their fraction in  those with  $\Delta r_i>0.3$  
is about $ 0.2$. 
This means that the distinct part 
 is negligible as compared to the self part in  
${\cal{F}}_i$ in Eq.(4.2), in accord with the  calculation 
by La\v{ c}evi\'{c} {\it et al.}\cite{Lacevic}.
(iv) 
Among  $366$  particles with $\Delta r_i>0.3$  
and ${\cal F}_i=1$, most of them (339) are $\bi B$ particles.

\begin{table}
\caption{Particle numbers in Figs.15 and 16  
at $t=10000$ and 10200 for 
$N=10^4$ in 3D.}
\begin{tabular}{|c|c|c|c|c|} 
\hline
\hspace{-3mm}$t$ & 
${\cal B}>0~~$ & ${\cal B}>0$,${\cal F}=0$  & 
${\cal B}={\cal F}=0$&$\Delta r>0.3$ \\
\hline
$10000$ & 2351 & 641 & 811 & 1818\\
\hline
$10200$  & 2393 & 675 & 898 & 1943 \\
\hline
common  & 2094 & 400 & 268&980  \\
\hline
\end{tabular}
\end{table}

\begin{table}
\caption{Numbers of non-$\bi B$ and $\bi B$ particles 
for three categories at $t=10000$ for 
$N=10^4$ in 3D  in Figs.13 and 14.   }
\begin{tabular}{|c|c|c|c|c|} 
\hline
\hspace{-3mm} 
& 
  ${\Delta r}>0.3$  & 
${\Delta r}>0.3$& ${\Delta r}<0.3$ &  \\
& 
   $ {\cal F}=0$ & 
$ {\cal F}=1$& ${\cal F}=1$& total \\
\hline
${\cal B}=0$ & 811 & 27 &  6811&7649 \\
\hline
${\cal B}>0$&641  & 339 &  1371 & 2351   \\
\hline
total & 1452 & 366 & 8182& 10000 \\
\hline
\end{tabular}
\end{table}

\section{Summary and remarks}
\setcounter{equation}{0}

We have examined the dynamic heterogeneity of glassy particle systems 
in the bond-breakage scheme \cite{yo,yo1} and  in the four-point scheme 
\cite{Lacevic}. The former treats  
 the irreversible configuration changes, 
while in  the latter also included are  
 the  reversible particle displacements  due to 
the low-frequency vibrational modes. 
These two kinds of  motions are both highly 
heterogeneous in glassy states. 

Our main  results are as follows.\\
(i)  In Sec.III,  we have generalized the 
 bond breakage theory \cite{yo,yo1} to define 
the   broken bond number 
 ${\cal  B}_i(t_0,t_1)$, the fractions of the particles with $k$ broken 
bonds $\phi_b(t,k)$,  the  correlation function $G_b(r,t)$, 
the structure factor $S_b(q,t)$, and 
the susceptibility  $\chi_b(t)$ in Eqs.(3.12), (3.14), and (3.19)-(3.21). 
We have defined the bond-breakage time $\tau_b$  
 in Eq.(3.4) and   the bond-preserving time $\tau_{bp}$ in Eq.(3.16) 
in addition to  the relaxation time 
$\tau_\alpha$ from $F_s(q,t)$ in Eq.(3.6). 
In Fig.3, $\chi_b(t)$ exhibits a maximum as a function of $t$,  yielding the 
maximization time $t_b^{\rm max}$, 
while   the Ornstein-Zernike fitting of  
 $S_b(q,t_b^{\rm max})$  yields $\xi_b=\xi_b(t_b^{\rm max})$ in Fig.4.  
These quantities are nearly independent of the system size 
as long as $1\ll \xi_b \ll L$. \\
(ii)  In Sec.IV, we have discussed the four-point theory, 
where  the  overlap function $w(r)$ in Eq.(4.3)    defines   
the initial circles (spheres) in 2D (3D).  
The  overlap  number ${\cal F}_i(t_0,t_1)$ 
in Eq.(4.2)   determines  
the  correlation function $G_4(r,t)$, 
the   structure factor $S_4(q,t)$, and 
the   susceptibility   $\chi_4(t)$ in Eqs.(4.12)-(4.14). 
Maximization    of   $\chi_4(t)$ with respect  to $t$  yields 
the characteristic time $t_4^{\rm max}$.
We have shown that the nonoverlap motions 
 from the initial circles stem from the 
 thermal  excitation  of the low-frequency vibrational modes 
 and  the escape jumps from temporary cages  as in Figs.5, 6, 11, and 13. 
The thermal collective motions   appear  from the  initial stage  
($t\gs 10)$ as  in Fig.7, 
while the jump motions  emerge very  slowly.   
The maximization procedure 
 of   $\chi_4(t)$ with respect to the overlap length 
 \cite{Lacevic} is  to maximize 
 the contribution from 
the thermal collective motions to  $\chi_4(t)$. 
 In 2D,  the system-size dependence of 
the four-point correlations is strong  at long wavelengths 
even for $\xi_4\ll L$.\\ 
(iii) In Sec.V, we  have compared  the relaxation times 
$\tau_b$, $\tau_{pb}$, $\tau_\alpha$, $t_b^{\rm max}$, and  
$t_4^{\rm max}$ in Fig.12 to obtain the sequence (5.1), 
where $\tau_\alpha (\sim t_4^{\rm max})$ 
is considerably shorter than $\tau_b$. 
Next we have presented  snapshots of the displacements, 
${\cal  B}_i$,  and ${\cal F}_i$ at $t=10^4$ 
for $N= 4000$ with marked large-scale heterogeneities in Fig.13.  
We  have grouped  the particles with ${\cal F}_i=0$ 
into ${\bi B}$ and non-${\bi B}$ particles, 
where they are those with and without 
broken bonds, respectively. The patterns of the ${\bi  B}$ 
particles with ${\cal F}_i=0$ closely resemble those 
of the total ${\bi  B}$ particles. 
The non-$\bi B$ particles with ${\cal F}_i=0$  arise 
from the low-frequency   vibration modes undergoing 
relatively rapid temporal variations, as can be seen 
in the inset of Fig.6 and in  Fig.14.\\
(iv)  Also in 3D,  the four-point correlations 
arise from the  thermal 
 collective motions with ${\cal B}_i=0$ 
 and the bond-breakage   motions with ${\cal B}_i\ge 1$  
as in Figs.15 and 16. As a charcteristic feature in 3D, 
 Table IV shows that about $60\%$ of the   $\bi B$ paricles satisfy  
$\Delta r_i<0.3$ and ${\cal F}_i=1$. These 
particles suround the particles 
which have undergone string-like  motions. \\

We  make some remarks  in the following.
\\ 
(1) 
The heterogeneity exhibited by   the low-frequency vibration modes 
still remains largely 
 unexplored \cite{Ruocco,Shintani,Liu,Reichman,Barrat,Bonn,Br}.
In future work, we should examine  how  it depends on 
 the size ratio   
and the composition \cite{Hama,Jack,Shiba}.  
  \\ 
(2)   
The vibration modes  determine  
 the plateaus  of the time-correlation function 
$F_s(q,t)$ in Eq.(3.5) and the mean square displacement 
$M(t)$ in Eq.(A4), resulting in significant 
 system-size effects  \cite{finite,Kim}. 
They also give rise to the system-size dependence of 
the four-point structure $S_4(q,t)$ at small $q$ 
in  (b) of Fig.10.    
Our present analysis  is mostly for 2D, but 
Eqs.(A2) and (A8) provide one possible sourse of the finite size effect 
in  3D. \\ 
(3) We also comment  on the effect of 
a thermostat, which 
 was used only in preparing the initial states.  
We have  found that a thermostat 
can strongly affect   the low-frequency  vibration 
modes (not shown in this paper). 
For example, the second  peak of $\chi_4(t) $ 
 at $t \cong L/2c_\perp$ for  $T\le 0.80$ in Fig.9 disappeared  
in the presence of a thermostat, presumably because 
it effectively  increases the acoustic damping. \\ 


\begin{acknowledgments}
This work was supported by Grant-in-Aid 
for Scientific Research  from the Ministry of Education, 
Culture,  Sports, Science and Technology of Japan.  
T. K. was supported by the Japan Society for Promotion of Science.
The authors would like to thank   
Ryoichi Yamamoto,  Kunimasa  Miyazaki, Shinichi Sasa, 
and Kang Kim  for 
informative discussions. 
The numerical 
calculations were carried out on SR16000 at YITP in Kyoto
University and 
Altix ICE 8400EX at ISSP in University of Tokyo. 
\end{acknowledgments}

\vskip5mm
\hspace{-0.3cm}
{\bf Appendix: Thermal positional fluctuations and finite size effect }\\
\setcounter{equation}{0}
\renewcommand{\theequation}{A\arabic{equation}}

In  solids, 
the vibration modes give rise to the thermal fluctuations of the particle 
displacements ${\bi u}_i(t)= {\bi r}_i(t) -{\bar{\bi r}}_i$ with 
${\bar{\bi r}}_i$   being  the time-averaged  positions.  
In our theory, we may define   
${\bar{\bi r}}_i$  on timescales shorter than the bond-preserving time 
$\tau_{\rm bp}$ in Eq.(3.16). 
The contributions from  the large-scale  modes  
may be calculated using   the classical linear elasticity theory, 
which should be  valid at  sufficiently 
long wavelengths 
even in glass \cite{Barrat}.

In 2D,  we may  expresss   the displacement variance 
$\av{|{\bi u}|^2}=\sum_i \av{|{\bi u}_i|^2}/N$  
 as in Eqs.(4.6) and (4.7). 
For  2D solids \cite{Janco},  
use   has been made of  the relation,   
\be 
\av{|{\bi u}_i(t)-{\bi u}_j(t)|^2}= 2 C_1 \ln (r_{ij}/a_0), 
\en  
where $C_1$ is given in E.(4.7),   $r_{ij}$ 
is the distance between particles $i$ and $j$,  and $a_0$ is a 
microscopic length.  Thus,  Eq.(A1) represents the anomalous long-range 
correlation in 2D solids. 
These  expressions folllow if 
the discrete sums over the long-wavelength  vibration 
modes ($\sum_{\bi q} |{\bi q}|^{-2}$) 
are  replaced by the  wave-number integral ($ 
\int dq q^{d-3}$).   
In 3D,  there is no long-wavelength  divergence, but  
 the lower bound of the wave number ($\propto L^{-1}$) yields  
the following $L$ dependence, 
\be 
\av{|{\bi u}|^2}=   D_0 -  D_1 /L, 
\en 
where $D_0$ and $D_1$ are functions of  $T$. 
If we perform  the corresponding discrete summation   
over the  modes 
under the periodic boundary condition, we  obtain 
\be 
D_1= 0.21 {k_BT}\bigg(\frac{2}{\mu}+ 
\frac{1}{K+4\mu/3}\bigg). 
\en

For $t \ll \tau_{bp}$, 
we may  set  ${\bi r}_i(t_0)-{\bi r}_i(t_0+t)
= {\bi u}_i(t_0)-{\bi u}_i(t_0+t)$ by 
 neglecting  the configuration changes.  
The mean square displacement $M(t)$ is written as 
\bea 
M(t) &=& 
\sum_i \av{|{\bi r}_i(t_0)-{\bi r}_i(t_0+t)|^2}/N\nonumber \\
&\cong &  \sum_i \av{|{\bi u}_i(t_0)-{\bi u}_i(t_0+t)|^2}/N.  
\ena 
If the cross correlation 
$\sum_i \av{{\bi u}_i (t_0+t)\cdot {\bi u}_i (t_0)}/N$ decays  to zero 
due to 
the acoustic damping 
before the $\alpha$  relaxation,    
a well-defined pleteau $M_{\rm p}$  appears in $M(t)$ with    
\be 
M(t)\cong M_{\rm p}= 
2\av{|{\bi u}|^2}, 
\en   
for $1\ll t \ll \tau_{bp}$
as already given in Eq.(4.6).
 
The time correlation function 
$F_s(q,t)$ in Eq.(3.5)  also assumes  a 
 well-defined   plateau  value  $f_{\rm p}=f_{\rm p}(T,N)  $ 
 at low $T$.  For the Gaussian distribution of ${\bi u}_i$, we have 
\cite{Hansen} 
\be 
f_{\rm p} \cong \exp[-q^2 M_{\rm p}/2d]= \exp[-q^2\av{|{\bi u}|^2}/d]  , 
\en 
If the structural relaxation time  $\tau_\alpha$ is defined by  Eq.(3.6), 
its  dependence on  $N (\propto N^{1/d})$  
can arise from  Eq.(4.6) in 2D and 
from Eq.(A2) in 3D.

For  2D, $f_{\rm p}$ depends on $N$ as  
\be 
f_{\rm p}\propto L^{-q^2C_1/2}\propto N^{-q^2C_1/4}.
\en     
For   $ T=0.56$  and  $q=2\pi$, 
our numerical analysis gives   $f_{\rm p}=  0.6 $ 
    for $N=4000$ in Fig.1    
and  $f_{\rm p}=  0.5 $ 
    for     $N=64000$. 
The ratio  of these two values $0.6/0.5=1.2$ 
is close to  the  theoretical 
ratio $16^{q^2C_1/4}= 1.17$ from Eq.(A7).  

 For 3D,  Eqs.(A2), (A5), and (A6) yield  
\be 
f_{\rm p }(T,N)/ f_{\rm p }(T,\infty) =   \exp[B_f  N^{-1/3}].  
\en 
Under  the periodic boundary condition, Eq.(A3) gives      
\be 
B_f \cong 0.14 q^2k_BT n^{1/3}/\mu,
\en  
where  $K$ is  assumed to be considerably larger than $\mu$.  
Kim and Yamamoto  \cite{Kim} studied the finite size effect 
 using  the soft-core potential  
for   $N=108, 10^3$, and $10^4$ in 3D.   
Their data of $F_s(q,t)$ at  $t=10$ for  $q=2\pi$ and  $T=0.267$    
may be approximately fitted to  the form 
$\propto \exp[ 0.25 N^{-1/3}]$, while  
 Eqs.(A8) and (A9) give 
$f_{\rm p } \propto \exp[0.29 N^{-1/3}]$ (where 
 we obtain $\mu=5$  
from the stress-strain relation).


\begin{references}
\bibitem{sound} In our simulation, 
the  transverse and longitudinal 
sound velocities are  given  by  
$c_\perp= ({\mu/\rho})^{1/2}\cong 4.1  
$ and $
c_\parallel= [(K+\mu)/\rho]^{1/2}\cong 8.9$  in units of $\sigma_1/\tau_0$ 
at $T=0.56$ in 2D, where   
 $\rho=1.06m_1\sigma_1^{-2}$ is   the average mass density.
\end{references}
\end{document}